\documentclass[english, 12pt]{article}

\usepackage{amsthm}\usepackage{physics}
\usepackage{gensymb}
\usepackage{flexisym}
\usepackage{tikz}
\usepackage{mathrsfs}\usepackage{rotating}\usepackage{titlesec}
\usepackage{lipsum}\usepackage{fancyhdr}
\usepackage{framed}\usepackage{chngcntr}\usepackage{blindtext}
\usepackage{epstopdf}
\usepackage{multirow}
\usepackage{colortbl}
\usepackage{pdflscape}
\usepackage{bbm}
\usepackage{xcolor}

\usepackage{float}
\usepackage{color}
\usepackage{textcomp}
\usepackage{mathrsfs}
\usepackage{amsmath}
\usepackage{amsthm}
\usepackage{amssymb,bm}
\usepackage{undertilde}
\usepackage{graphicx}
\usepackage{setspace}
\usepackage[authoryear]{natbib}
\usepackage{url}
\providecommand{\tabularnewline}{\\}

\makeatletter

\newcommand{\blind}{1}
\usepackage{geometry}
\usepackage{setspace}
\usepackage{standalone}

\usepackage[utf8]{inputenc}

\usepackage{epsfig}
\usepackage{amsfonts}
\usepackage{bbm}

\@ifundefined{showcaptionsetup}{}{%
 \PassOptionsToPackage{caption=false}{subfig}}
\usepackage{subfig}
\makeatother

\newcommand{\bbeta}{\boldsymbol{\beta}}

\newcommand{\bsigma}{\bm{\sigma}}

\newcommand{\bTheta}{\bm{\Theta}}
\newcommand{\bphi}{\bm{\phi}}

\newcommand{\btau}{\bm{\tau}}

\newcommand{\bZ}{\textbf{Z}}
\newcommand{\bs}{\bm{s}}

\newcommand{\Cov}{\textbf{Cov}}

\newcommand{\bH}{\textbf{H}}

\newcommand{\by}{\bm{y}}

\newcommand{\bV}{\textbf{V}}
\newcommand{\bW}{\bm{W}}
\newcommand{\bD}{\textbf{D}}

\newcommand{\btheta}{\bm{\theta}}

\newcommand{\bx}{\textbf{x}}

\newcommand{\bO}{\mathbf{0}}

\newcommand{\bz}{\bm{z}}

\newcommand{\bI}{\textbf{I}}

\newcommand{\calN}{{\cal N}}

\newcommand{\cov}{\operatorname{cov}}

\makeatother

\usepackage{authblk}
\defcitealias{NASA}{NASA2020}

\expandafter\def\expandafter\normalsize\expandafter{%
    \normalsize
    \setlength\abovedisplayskip{6pt}
    \setlength\belowdisplayskip{6pt}
    \setlength\abovedisplayshortskip{6pt}
    \setlength\belowdisplayshortskip{6pt}
}

\allowdisplaybreaks

\usepackage{babel}

\begin{document}

\begin{singlespace}

\if1\blind {
\title{Bayesian Latent Variable Co-kriging Model in Remote Sensing for Observations with Quality Flagged}
 \author[1]{Bledar A. Konomi \thanks{Corresponding author:
Bledar A. Konomi (alex.konomi@uc.edu)} } 
 \author[1]{Emily L. Kang}
  \author[1]{Ayat Almomani}
\author[2]{Jonathan Hobbs}
\affil[1]{Division of Statistics and Data Science, Department of Mathematical Sciences, University of Cincinnati}
\affil[2]{Jet Propulsion Laboratory, California Institute of Technology}
\maketitle 
} \fi
\end{singlespace}
\begin{abstract}
\begin{singlespace}

Remote sensing data products often include quality flags that inform users whether the associated observations are of good, acceptable or unreliable qualities. However, such information on data fidelity is not considered in remote sensing data analyses. Motivated by observations from the Atmospheric Infrared Sounder (AIRS) instrument on board NASA's Aqua satellite, we propose a  latent variable co-kriging model with separable Gaussian processes to analyze large  quality-flagged remote sensing data sets together with their associated quality information. We augment the posterior distribution by an imputation mechanism to decompose large covariance matrices into separate computationally efficient components taking advantage of their input structure. Within the augmented posterior, we develop a Markov chain Monte Carlo (MCMC) procedure that mostly consists of direct simulations from conditional distributions. In addition, we propose a computationally efficient recursive prediction procedure. We apply the proposed method to air temperature data from the AIRS instrument. We show that incorporating quality flag information in our  proposed model substantially improves the prediction performance compared to models that do not account for quality flags. 
\end{singlespace}
\end{abstract}
\begin{singlespace}
Keywords:  Co-kriging; Gaussian process;   Markov chain Monte Carlo; remote sensing; separable covariance function.
\end{singlespace}


\newpage

\section{Introduction}
Remote sensing technology provides a wealth of information for understanding geophysical processes with unprecedented spatial and temporal coverage. Remote sensing data provide indirect information on these geophysical quantities of interest, which are typically estimated or inferred from instrument spectra \citep{Susskind2003}. The heterogeneous nature of the Earth's atmosphere and surface contributes to remote sensing data records with variable quality, which is often documented by the product development teams and included as quality flags associated with observations in remote sensing data products \citep[e.g.,][]{AIRSV7Guide}. These quality flags, instead of providing quantitative uncertainty measures for the observations, indicate whether data are of good, acceptable or unreliable qualities. When the remote sensing data products are used in a wide range of downstream analyses, these quality flags are usually treated in a dichotomous way: Data flagged as unreliable are removed, while the remaining data, no matter how they may be flagged differently (e.g., good or acceptable), are combined directly \citep[e.g.,][]{Zhu2015,MaKangFusion2020,AIRSV7Lev3,waliser_obs4mips}. This practice ignores the delicate quality difference between observations with different quality flags.

Our work is motivated by data products from the Atmospheric Infrared Sounder (AIRS) instrument on board NASA's Aqua satellite. The AIRS instrument collects radiance spectra, termed Level 1 data products, across the globe in the infrared portion of the spectrum. A retrieval algorithm is used to infer atmospheric quantities such as temperature and humidity from the spectra at 45km$\times$ 45km spatial resolution, called the Level 2 data products \citep{Susskind2003}. Note that the AIRS Level 2 data products include a quality flag (QF) variable: QF value of $0$ indicates observations of very good quality, QF value of  $1$ indicates observations of acceptable quality, and  QF value of $2$ are deemed to be bad or unreliable. 
 AIRS data products have been used by weather prediction centers around the world to improve weather forecasts \citep{IRAssimilate}. They are also used to assess the skill of climate models and in applications ranging from volcanic plume detection \citep{volcanoairs} to drought forecasting \citep{droughtrmsns}. Although data with QF value of 0 are removed from such analyses, AIRS data with QF values of 1 and 0 are combined without further consideration of their quality difference indicated by their QF values \citep[e.g.,][]{AIRSV7Lev3,waliser_obs4mips}. In this work, we focus on ARIS Level 2 air temperature data at 100 different vertical pressure levels which we refer to as 3D (longitude, latitude, and vertical atmospheric pressure level) air temperature observations. To our knowledge, our work is the first to build a statistical model for such quality-flagged remote sensing data.

Since the quality flags indicate variable fidelity associated with remote sensing observations, a natural statistical model to explore the dependencies of successive fidelity levels is the autoregressive co-kriging model \citep{kennedy2000predicting}.  Essentially, this autoregressive co-kriging method considers a scalar and an additive discrepancy to model observations with a sequential fidelity order.  A number of important variations of this model have been proposed. \citet{QianWu2008} consider the scale discrepancy as a function of the input space by casting it as a Gaussian process (GP)  which produces non-standard conditional posteriors. \citet{perdikaris2017nonlinear} relax the auto-regressive structure by using deep learning ideas to introduce nonlinear relationships, but the computational cost to train the model is significantly increased. \citet{konomikaragiannisABTCK2019} propose a Bayesian augmented hierarchical co-kriging procedure which makes possible the analysis of non-nested input and non-stationary output.   These methods rely on Gaussian processes and do not scale well with large data sets. Meanwhile,  these methods are initially designed for deterministic computer models and thus don't include a term for measurement errors which is highly recommended in spatial statistics \citep{Cressie1993,Stein99}.  


Spatial statistical methods for big data have been evolving over the past two decades due to emergence of massive spatial data sets. \cite{Banerjee2017} and \cite{heaton2017methods} provide  a good  overview  of  these methods with comparisons. Briefly we distinguish between low-rank approximation methods \citep{banerjee2008gaussian,cressie2008fixed}, approximate likelihood methods \citep{stein2004approximating,Gramacy2015}, sparse structures \citep{lindgren2011explicit,nychka2015multiresolution, MaKang2020,MeshedGP2020}, multiple-scale approximation \citep{Sang2012,Katzfuss2016}, and lower dimensional conditional distributions \citep{vecchia1988estimation,stein2004approximating,datta2016hierarchical,katzfuss2017general}. Directly applying these methods within the autoregressive co-kriging framework for multi-fidelity remote sensing data is complicated. In particular, when the locations of data at different fidelity levels are not nested, the likelihood function no longer benefits from the Markovian property assumed in the model,  making it complex to apply  the methods aforementioned for  massive spatial data \citep{konomikaragiannisABTCK2019}. Recently, \cite{cheng2020hierarchical} extend the nearest-neighbor Gaussian process (NNGP) to analyze multi-level data sets by introducing a nested NNGP reference set for each level. Although this method can handle multi-fidelity data, its way to define the reference sets makes it more appropriate when low-fidelity data are observed homogeneously over the spatial domain, but not suitable for the AIRS data described in Section~\ref{sec: data} when the observations with the same QF values are clustered spatially.

In this paper, we propose a latent variable co-kriging model with separable Gaussian processes, which is able to account for multi-fidelity remote sensing data with measurement errors. Noticing that for AIRS Level 2 data products the latitude and longitude of observation locations are fixed across different vertical pressure levels, we adopt a multiplicative (separable) covariance function for the horizontal and vertical dimensions. Such a multiplicative (separable)  covariance function is widely used in spatial statistics \citep{BanerjeeGelfand2014} and uncertainty quantification for computer experiments \citep{gramacy2020surrogates}. For AIRS air temperature data, the resulting covariance matrix can be decomposed into a Kronecker product of a purely horizontal (i.e., latitude and longitude) 
correlation matrix and a purely vertical (i.e., pressure level) correlation matrix.  This can alleviate the computational bottleneck related to Gaussian process likelihood evaluation and spatial prediction \citep{genton2007,Rougier2008,BilionisJCP2013,  Guillas2018, Ma_Konomi_Kang2019Env}.  However, the introduction of the different QF values as well as missing data destroys this Kronecker-product representation of the separable covariance structure. To facilitate efficient inference,  we introduce an imputation mechanism within the Markov chain Monte Carlo (MCMC) procedure to take advantage of the latent variable representation and the data structure, which enables us to decompose the large covariance matrices into two separate computationally efficient components. Moreover, we propose a computationally efficient Monte Carlo recursive prediction procedure to make spatial prediction at high-fidelity level. We apply this proposed method to analyze level 2 AIRS air temperature data. Extensive numerical results demonstrate that compared to methods that ignores the different QF values of data, our method provides more accurate predictions but also remains computationally efficient.

The  rest  of  the  paper  is  organized  as  follows.   In  Section~\ref{sec: data}, we introduce the quality-flagged AIRS 3D air temperature data studied in this work. Section~\ref{sec: model} presents the latent variable co-kriging model. In Section~\ref{sec: infe}, we give details on Bayesian inference including spatial prediction by constructing an augmented posterior using imputation of latent variables. In Section~\ref{sec: res}, we apply the proposed method to analyze the AIRS air temperature data and demonstrate its inferential advantages compared to methods ignoring the quality flags. We offer conclusions and discussion in Section~\ref{sec: conc}.

\section{Description of the AIRS Air Temperature Data}\label{sec: data}

The Atmospheric Infrared Sounder (AIRS) instrument on board NASA's Aqua satellite measures radiance spectra in infrared channels along the satellite's polar orbit. These infrared channels are sensitive to thermal emission from the atmosphere. The AIRS retrieval algorithm first obtains cloud properties and effective cloud-cleared radiance (CCR) and then uses CCR to further infer atmospheric properties including temperature and humidity at different vertical pressure levels at 45km $\times$ 45km spatial resolution. A single 45km $\times$ 45km areal unit is known as a field of regard (FOR) in \cite{Susskind2003}. AIRS Level 2 data are further sectioned into pieces called granules. Each granule is roughly 2250km $\times$ 1650km and thus contains $n_s=1350$ FORs horizontally at each pressure level. Tere are a total of $n_p=100$ pressure levels in AIRS data products. Therefore, Within a granule, there are potentially $1,350\times 100=135,000$ in the three-dimensional domain (horizontally and vertically).

It has been noted that heterogeneity of clouds yields substantial variability in the radiance, resulting in varying quality of the retrievals. Therefore, most retrieved remote sensing data for atmospheric properties including air temperature are accompanied by a quality flag (QF) variable. The AIRS convention is that QF values of 0 indicate best quality data, mostly retrieved under a clear-sky condition, values of 1 indicate acceptable quality, and values of 2 are deemed bad quality. The  AIRS Level 2 products we analyze in this paper are extracted from the AIRS Version 6 retrieval support product \citep{Kahn2014}. Further details on the data products have been recently documented for the AIRS Version 7 products \citep{AIRSV7Guide}.

\begin{figure}[htb!]
	\centering
	\subfloat[\label{fig:da}]{\includegraphics[width=0.45\textwidth]{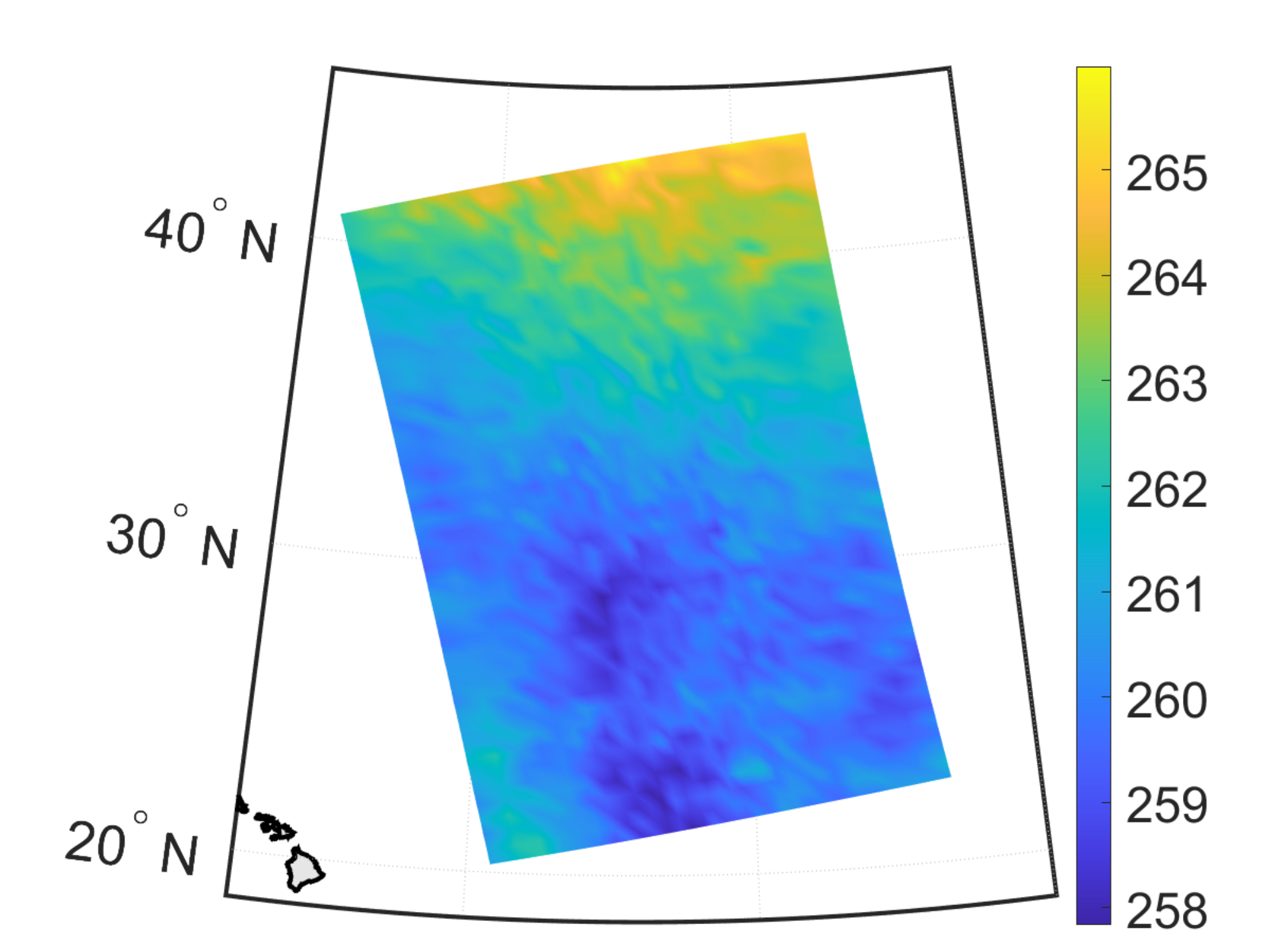}}
	\subfloat[\label{fig:db}]{\includegraphics[width=0.45\textwidth]{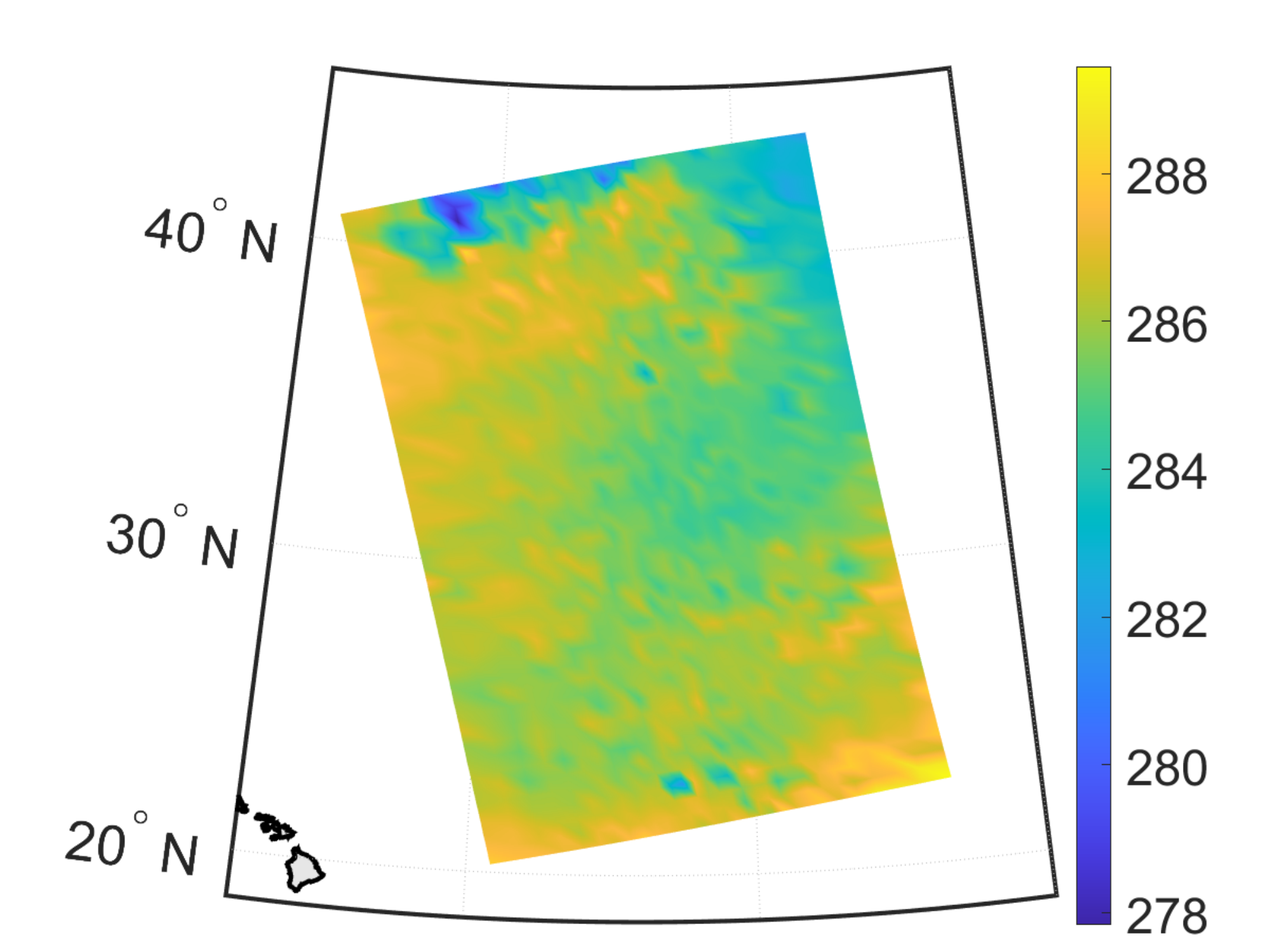}}\\
	\subfloat[\label{fig:dc}]{\includegraphics[width=0.45\textwidth]{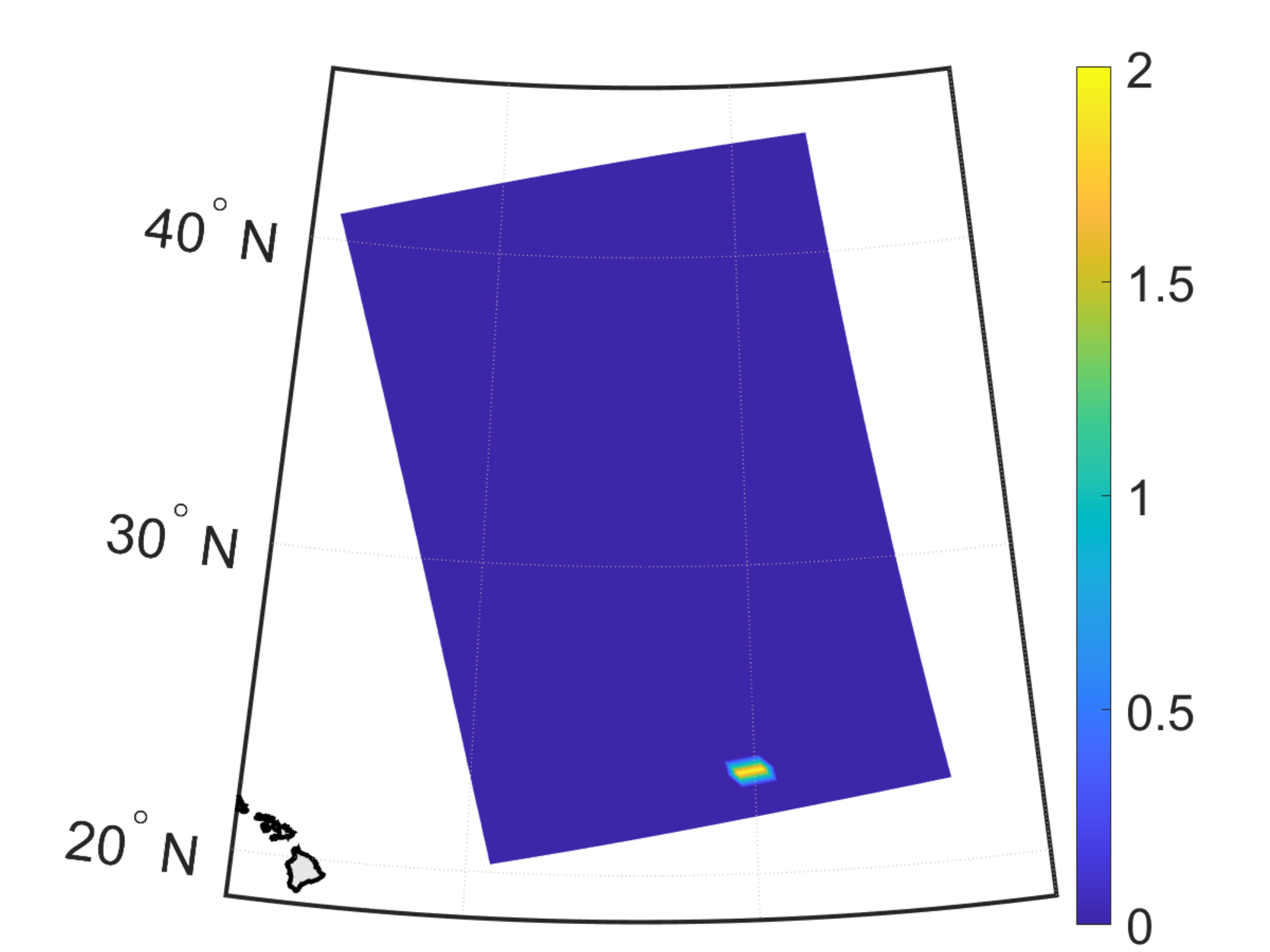}}
	\subfloat[\label{fig:dd}]{\includegraphics[width=0.45\textwidth]{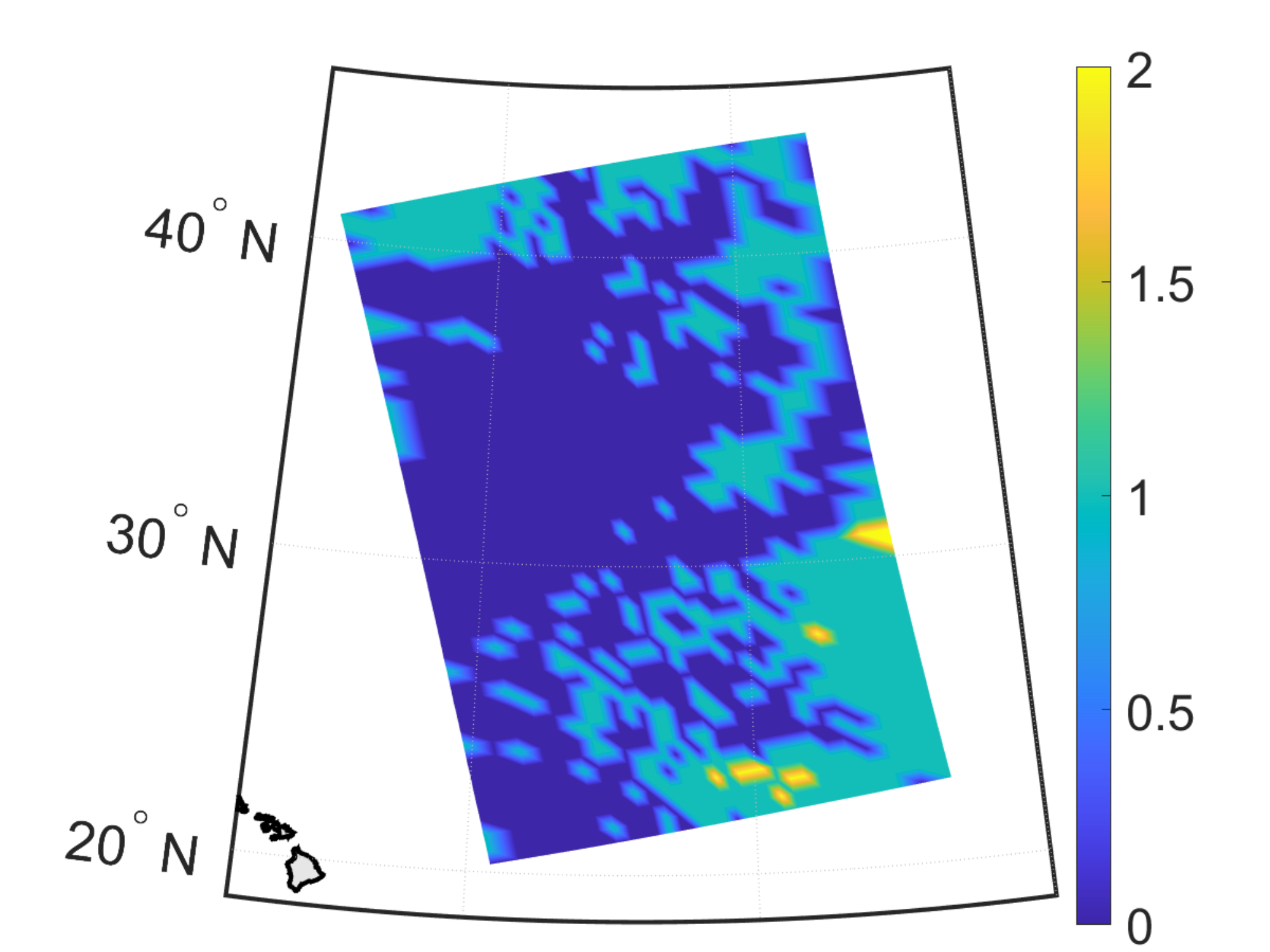}}
	\caption{Top: AIRS retrieved temperature (units are Kelvin) on two different pressure levels, a) pressure level 10 ($P=1.2972 \textrm{ hPa}$) and b) pressure level 90 ($P= 958.5911 \textrm{ hPa}$). Bottom: Retrieved temperature quality flag on two different pressure levels, c) pressure level  10 ($P=1.2972 \textrm{ hPa}$) and d) pressure level 90 ($P= 958.5911 \textrm{ hPa}$)}
		\label{fig:data4}
\end{figure}

In this paper, we focus on the 3D (longitude, latitude, and pressure level) Level 2 air temperature data product from AIRS, and we choose to focus on a specific granule over the subtropical eastern Pacific Ocean. This region's weather variability spans multiple cloud regimes that present multiple challenges for remote sensing. Therefore, the region corresponding to this particular granule has been chosen to be the study region in works assessing retrievals from AIRS and other instruments such as the MAGIC validation campaign \citep{ZhouKollias2015,Kalmus_MAGIC}. Figure \ref{fig:data4} displays the air temperature in our study region at two different pressure levels, together with their associated QF values. Note that here higher pressure levels means higher pressure values and thus closer to the surface, while lower pressure levels correspond to higher altitudes. The amount of observations with QF value of 1 (i.e., acceptable quality) and 2 (i.e., bad quality) increases as we move from low to high pressure levels. Panels (c) and (d) in Figure~\ref{fig:data4} show that there are a lot more observations with QF value 1 and 2 at pressure level 90 (pressure level value  $P= 958.5911 \textrm{ hPa}$) compared to those at pressure level 10 (pressure level value $P=1.2972 \textrm{ hPa}$).  Meanwhile, it is clear that observations with the same QF values form spatial clusters instead of being distributed uniformly in the region, which is expected as lower quality observations don't occur randomly but are related to heterogeneous cloud properties at medium or fine spatial scales. Because observations with QF value of 2 are not reliable data, we treat them as missing values in our analysis. For simplicity, we will refer to  air temperature data with QF value of 1 as low-fidelity observations and those with QF value of 0 as high-fidelity observations.

\section{The Latent Variable Co-kriging Model}\label{sec: model} 

Let $ \textbf{x} =(\textbf{s},p) \in \mathcal{X}  = \mathcal{S} \times \mathcal{P}$ represent the input space with the $\textbf{s}=(s_1,s_2) \in \mathcal{S}\subset \mathbb{R}^2 $ denoting the longitude and latitude of the FOR centers and $p \in \mathcal{P} \subset \mathbb{R} $ denoting the pressure level values. Let  $ \left\lbrace (z_{L}(\textbf{x}), z_{H}(\textbf{x})): \textbf{x} \in \mathcal{X}\right\rbrace $ represent the observation of low fidelity (QF value  1, subscript $L$) and high fidelity (QF value 0, subscript $H$).    We assume the data observed at input $\bx$ is  contaminated by additive random noise: $z_{L}(\textbf{x})=y_{L}(\bx)+\epsilon_{L}(\bx)$ and  $z_{H}(\textbf{x})=y_{H}(\bx)+\epsilon_{H}(\bx)$.
Specifically, we model the measurement error of the low fidelity as $\epsilon_{L}(\bx)\sim N(0,\tau_{L}^2)$ and the measurement error of the high fidelity as $\epsilon_{H}(\bx)\sim N(0,\tau_{H}^2)$.


For the latent low- and high-fidelity processes $y_{L}(\bx)$ and  $y_{H}(\bx)$, we assume that they are linked via the latent variable autoregressive co-kriging model as follows:
\begin{align}
 y_{L}(\bx)   &= \mathbf{h}_{L}(\bx)^T\boldsymbol{\beta}_{L}+w_{L}(\bx),\nonumber \\
 y_{H}(\bx) &=  \rho y_{L}(\bx)+\delta_{H}(\bx)\\
 &=  \rho y_{L}(\bx)+\mathbf{h}_{H}(\bx)^T\boldsymbol{\beta}_{H}+w_{H}(\bx), \nonumber\label{eq:davdsgdaf}
\end{align}
where $\rho$ represents a constant scalar discrepancy, and $\delta_{H}(\bx)$ represents the additive discrepancy between $y_{L}(\bx)$ and  $y_{H}(\bx)$. Here,  $\mathbf{h}_{L}(\cdot)$ and $\mathbf{h}_{H}(\cdot)$ are $p_H$ and $p_L$ known covariates to describe the mean structure of $y_{L}(\bx)$ and  $\delta_{H}(\bx)$, respectively. Utilizing the structure of the observations,  we use a  product of separate basis functions for the spatial locations and the pressure level values. Suppose, we wish to use $m_{L,p}$ basis functions to capture the pressure dependence of the lower fidelity mean: $h_{L,p}=\{h_{L,p,1}(p),\dots, h_{L,p,m_{L,p}}(p) \}$. We choose also $m_{L,s}$ basis functions to capture the spatial dependence of the mean: $h_{L,\bs}=\{h_{L,\bs,1}(\bs),\dots, h_{L,\bs,m_{L,s}}(\bs) \}$. These can be for example the finite element basis of the model or any other suitable functions.  This strategy makes possible to separately construct basis functions, in lower dimensions,  with the help of the Kronecker product as we explain below.  Correspondingly, $\boldsymbol{\beta}_{L}$ and $\boldsymbol{\beta}_{H}$ are the unknown vectors of coefficient vectors. Finally, we  model the latent  $w_{H}(\cdot)$ and $w_{L}(\cdot)$ as mutually independent Gaussian processes:  $w_{L}(\cdot)\sim GP(0,\sigma_L^2R_{L}(\cdot,\cdot;\boldsymbol{\theta}_{L}))$ where $\sigma_L^2$ is the variance and $R_{L}(\cdot,\cdot;\boldsymbol{\theta}_{L})$ is a correlation
function with  parameters $\boldsymbol{\theta}_{L}$ at fidelity  level $L$. Similarly,  $w_{H}(\cdot)\sim GP(0,\sigma_H^2R_{H}(\cdot,\cdot,\boldsymbol{\theta}_{H}))$, but independent of $w_L(\cdot)$.  

We define $\bsigma^{\boldsymbol{2}}=(\sigma_L^2,\sigma_H^2)$, $\btau^{\boldsymbol{2}}=(\tau_L^2, \tau_H^2)$, and $\bbeta=(\bbeta_{L},\bbeta_{H})$, $\boldsymbol{\theta}=(\boldsymbol{\theta}_{L}, \boldsymbol{\theta}_{H})$. A direct Bayesian inference can be computationally costly when the number of the observations is large. The computational complexity of the likelihood is $\mathcal{O}(n_{s}^3n_{p}^3)$  flops and  the  parameters of the model cannot be expressed in  closed form.  For more details see  Appendix A.

To overcome this computational issue, we assume that the covariance function can be expressed in a multiplicative (separable) form $R_{L}(\textbf{x},\textbf{x}^{\prime};\boldsymbol{\theta}_{L}) = R_{L,s}(\textbf{s},\textbf{s}^{\prime}; \boldsymbol{\theta}_{L,s})  R_{L,p}(p,p^{\prime}; \boldsymbol{\theta}_{L,p})$ and $R_{H}(\textbf{x},\textbf{x}^{\prime};\boldsymbol{\theta}_{H}) = R_{H,s}(\textbf{s},\textbf{s}^{\prime}; \boldsymbol{\theta}_{H,s})  R_{H,p}(p,p^{\prime}; \boldsymbol{\theta}_{H,p})$.   This multiplicative (separable) model  requires the specification of correlation functions for space and atmospheric pressure for each of the two fidelity levels $L$ and $H$. To ensure the well-conditioning of the correlation matrices involved in the calculations, we add nugget effects for the spatial component and   for the atmospheric pressure component. In the literature these values are usually fixed based on a  cross validation exploration and they are expected to be typically small (of the order of $10^{-6}$). In practice, considering these parameters as unknown and estimating them from the data not only improves the stability of the computations but  also can lead to better predictive accuracy as shown in our application.  


Based on a preliminary study on AIRS data, for both levels of fidelity, we choose a product  exponential correlation function:  \begin{align}
R_{\cdot,s}(\bs,\bs'|\boldsymbol{\theta}_{\cdot,s}) & =\text{exp}\left(-\sum_{i=1}^{2}\frac{\left|s_{i}-s_{i}'\right|}{\phi_{\cdot,i,s}}\right)+g_{\cdot,s}^2\delta(\textbf{s}, \textbf{s}^{\prime}) \\
R_{\cdot,p}(p,p'|\boldsymbol{\theta}_{\cdot,p}) & =\text{exp}\left(-\frac{\left|p-p'\right|}{\phi_{\cdot,p}}\right)+g_{\cdot,p}^2\delta(p, p^{\prime}),
\end{align}
where $\boldsymbol{\theta}_{\cdot,s}=(\boldsymbol{\phi}_{\cdot,s},g_{\cdot,s}^2)$, $\boldsymbol{\theta}_{\cdot,p}=(\phi_{\cdot,p},g_{\cdot,p}^2)$, $\phi_{\cdot,1,s}$ controls the spatial dependence strength in longitude, $\phi_{\cdot,2,s}$ controls the spatial dependence strength in latitude, and $\phi_{\cdot,p}$ controls the spatial dependence strength in the pressure level values.  Also ${g}_{\cdot,s}^2$ accounts for the nugget effect in the spatial component and ${g}_{\cdot,p}^2$  accounts for the nugget effect in the atmospheric pressure component.  More intricate covariance functions  such as the Mat{\'e}rn family  \citep{Cressie1993, Stein99, BanerjeeGelfand2014} or non-stationary models \citep{Paciorek06,Konomi2014JCGS} can also be used within the proposed latent variable co-kriging model.  

Despite this flexible and general representation of the separable covariance function, the covariance matrix of the observations or the latent variables cannot be represented as a  Kronecker product of  separate components. This is simply because none of the input combinations of low and high fidelity data can be represented as a tensor product.  The presence of missing data for both quality flags destroys the Kronecker product representation of the covariance matrices and makes the computations impossible in practice.

\section{Bayesian Inference}\label{sec: infe}

We make  computations and practical implementation possible by constructing an augmented posterior  which is based on the imputation of latent variables. The augmented posterior takes advantage of the tensor product of the locations and the pressure level values in a granule.  Based on this augmented posterior, we are also able to construct a MCMC procedure with mostly closed form conditional distributions for parameter inference as well as a computationally efficient recursive prediction procedure.



\subsection{Augmented Posterior}\label{sec: aug}
Let $\{\bD^{H},\bD^{L},\bD^{M}\}$ be the observed input location set of QF $0$, QF $1$, and QF $2$ respectively. The union of these three input $\tilde{\bD}=\bD^{H}\cup\bD^{L}\cup\bD^{M}$ represents a granule and is defined in a grid format. In addition, assume the  $\{\boldsymbol{z}_{H},\boldsymbol{z}_{L},\boldsymbol{z}_{M}\}$  be the  observed output set of quality flag $0$, quality flag $1$, and quality flag $2$ respectively. Assume sets of points $\mathring{D}^{L}$ such that   $\mathring{\bD}^{L}=\bD^{H}\cup\bD^{M}$ is defined as the relative complement of $\bD^{L}$ in $\tilde{\bD}$. Further, assume sets of points $\mathring{D}^{H}$ such that   $\mathring{\bD}^{H}=\bD^{L}\cup\bD^{M}$ is defined as the relative complement of $\bD^{H}$ in $\tilde{\bD}$. 
 Let $\mathring{\boldsymbol{z}}_{L}=z_{L}(\mathring{D}^{L})$ and $\mathring{\boldsymbol{z}}_{H}=z_{H}(\mathring{D}^{H})$ be the missing output values of the temperature at the corresponding
input points in the low fidelity (quality flag $1$) and high fidelity  (quality flag $0$) data, respectively. We refer to $\{\mathring{\boldsymbol{z}}_{L},\mathring{\bD}^{L}\}$ and $\{\mathring{\boldsymbol{z}}_{H},\mathring{\bD}^{H}\}$ 
as the missing data set. We also refer to $\{\tilde{\boldsymbol{z}}_{L},\tilde{\bD}\}$ and $\{\tilde{\boldsymbol{z}}_{H},\tilde{\bD}\}$
as the complete data set of the low and high fidelity level respectively.

Assume that $\tilde{\textbf{w}}_L$ is the latent variables  of the low fidelity level (QF 1) obtained in $\tilde{\bD}$ and  $\tilde{\textbf{w}}_H$  is the latent variables of the high fidelity level (QF 0) obtained in  $\tilde{\bD}$. The covariance function of both $\tilde{\textbf{w}}_L$ and $\tilde{\textbf{w}}_H$ is defined in a grid where we can take advantage of the Kronecker product representation of the separable covariance function. For simplicity in the formulation below, let $\bTheta=(\boldsymbol{\theta},\bbeta,\rho,\bsigma^2,\btau^2)$, $\tilde{\textbf{w}}=(\tilde{\textbf{w}}_{L},\tilde{\textbf{w}}_{H})$, and $\tilde{\boldsymbol{z}}=(\tilde{\bz}_{L},\tilde{\bz}_{H})$.
We assign a prior distribution on the parameter $(\bTheta,\tilde{\textbf{w}})$, such as: 
\[p(\bTheta,\tilde{\textbf{w}}) =p(\bbeta_{L},\sigma^2_{L})p(\btheta_{L}) p(\rho,\bbeta_{H},\sigma^2_{H})p(\btheta_{H}) p(\tau_{L}^2)p(\tau_{H}^2)p(\tilde{\textbf{w}}_{L}|\btheta_{L})p(\tilde{\textbf{w}}_{H}|\btheta_{H}).\]
The GP parameters  are assumed a priori independent of each other for different fidelity level data. Within each fidelity level, we choose non-informative priors for  hyper-parameters $(\bbeta_{L}, \sigma_{L}^2)$ and $((\bbeta_{H},\rho), \sigma_{H}^2)$ and  $\boldsymbol{\phi}_{\cdot,s}$,$\boldsymbol{\phi}_{\cdot,p}$, $g_{\cdot,s}^2$, and $g_{\cdot,p}^2$ are considered to be independent truncated \textit{Gamma} distributions.

The joint posterior distribution of the above model given   observations $\boldsymbol{z}=(\bz_{L},\bz_{H})$ is:
\begin{align}\label{eq:phiposterior-1}
p(\bTheta,\tilde{\textbf{w}},\mathring{\boldsymbol{z}}| \bz)  & \propto  p(\mathring{\boldsymbol{z}}|\bTheta,\tilde{\textbf{w}}, \bz)  p(\boldsymbol{z}|\bTheta,\tilde{\textbf{w}})p(\tilde{\textbf{w}}|\btheta)p(\bTheta) \nonumber\\ &
= f(\tilde{\bz}_L|\btheta_{L},\bbeta_{L},\tau_{L}^2,\tilde{\textbf{w}}_{L}) f(\tilde{\bz}_H|\btheta_{H},\bbeta_{H},\rho,\tau_{H}^2, \tilde{\textbf{w}}_{H},\tilde{\textbf{w}}_{L})\\&
\times p(\bTheta_{L})p(\tilde{\textbf{w}}_{L}|\btheta_{L})p(\bTheta_{H})p(\tilde{\textbf{w}}_{H}|\btheta_{H}) \nonumber
\end{align}
where the conditional distributions $f(\tilde{\bz}_L|\cdot)$ and $f(\tilde{\bz}_H|\cdot)$ are  independent multivariate Gaussian probability density functions.  The joint posterior distribution of $\tilde{\textbf{w}}$ can be factorized in two parts, because the proposed augmentation artificially creates the same design for both latent variables. The Markovian condition induces the required conditional independence. Given the independent specification of the prior, it is easy to see that the augmented posterior is also factorized into two conditionally independent parts. Note that $p(\bTheta,\tilde{\textbf{w}},\mathring{\boldsymbol{z}}| \bz)$  admits  the  posterior  of  interest  $p(\bTheta,\textbf{w}| \bz)$  as  marginal  by  construction,  and  hence leads to the same Bayesian analysis. 

\subsection{MCMC Sampler}\label{Bayesian-inference}
We design a MCMC sampler, targeting the augmented posterior,
that involves a random permutation scan of blocks updating $[\mathring{\boldsymbol{z}}|\boldsymbol{z},\btheta,\boldsymbol{\sigma}^{\boldsymbol{2}},\boldsymbol{\beta}, \btau^{\boldsymbol{2}}]$, $[\tilde{\textbf{w}}|\tilde{\boldsymbol{z}},\btheta,\boldsymbol{\sigma}^{\boldsymbol{2}},\boldsymbol{\beta}, \btau^{\boldsymbol{2}}]$,  $[\boldsymbol{\beta},\boldsymbol{\sigma}^{\boldsymbol{2}},\btau^{\boldsymbol{2}}|\tilde{\boldsymbol{z}},\tilde{\textbf{w}},\btheta]$ from a direct Gibbs sampler with closed form distributions, and $[\btheta|\tilde{\boldsymbol{z}},\tilde{\boldsymbol{w}}, \boldsymbol{\beta},\boldsymbol{\sigma}^{\boldsymbol{2}},\btau^{\boldsymbol{2}}]$ from a Metropolis-Hastings (M-H) step. We are not only reducing computational complexity based on the covariance structure, but also are able to find closed form conditional distributions for $\sigma_{L}^2|\btheta_{L}, \tau_{L}^2,\tilde{\textbf{w}}$, $\sigma_{H}^2|\btheta_{H}, \tau_{H}^2,\tilde{\textbf{w}}$, $\tau_{L}^2|\sigma_{L}^2, \btheta_{L},\tilde{\bz},\tilde{\textbf{w}}$  and  $\tau_{H}^2|\sigma_{H}^2, \btheta_{H},\tilde{\bz},\tilde{\textbf{w}}$ and do not need  a high dimensional M-H that can slow down convergence.   Details regarding the MCMC blocks are explained below.

The full conditional posterior of $\tilde{\textbf{w}}_{L}$ is normal $\tilde{\textbf{w}}_{L} | \ldots   \sim N \bigg(\boldsymbol{\mu}_{\tilde{\textbf{w}}_{L}},  \boldsymbol{R}_{\tilde{\textbf{w}}_{L}}\bigg)$ and $\tilde{\textbf{w}}_{H}$ is normal  $\tilde{\textbf{w}}_{H} | \ldots   \sim N \bigg(\boldsymbol{\mu}_{\tilde{\textbf{w}}_{H}},  \boldsymbol{R}_{\tilde{\textbf{w}}_{H}}\bigg)$, where
\begin{align}
\boldsymbol{\mu}_{\tilde{\textbf{w}}_{L}}= &\boldsymbol{R}_{\tilde{\textbf{w}}_{L}}\Big[\frac{1}{\tau_{L}^{2}}\textbf{I}(\tilde{\textbf{z}}_{L}-\textbf{H}_{L} \beta_{L})+\frac{1}{\tau_{H}^{2}}\textbf{I}(\rho^{-1}(\tilde{\textbf{z}}_{H}-\textbf{H}_{H} \beta_{H}-\tilde{\textbf{w}}_{H})-\textbf{H}_{L} \beta_{L})\Big] \label{eq:etywsywet1}\\
\boldsymbol{R}_{\tilde{\textbf{w}}_{L}}= & \Big(\frac{1}{\tau_{L}^{2}}\textbf{I}+\frac{1}{\tau_{H}^{2}}\textbf{I} + \frac{1}{\sigma_{L}^{2}}  \boldsymbol{\textbf{R}_{L,s}^{-1} \otimes \textbf{R}_{L,p}^{-1}}  \Big)^{-1},\label{eq:etywsywet}\\
\boldsymbol{\mu}_{\tilde{\textbf{w}}_{H}}= & \boldsymbol{R}_{\tilde{\textbf{w}}_{H}} \frac{1}{\tau_{H}^{2}}\textbf{I}_{n} (\tilde{\textbf{z}}_{H}-\textbf{H}_{H} \beta_{H}-\rho(\textbf{H}_{L}\beta_{L}+\tilde{\textbf{w}}_{L})),\label{eq:etywsywet1}\\
\boldsymbol{R}_{\tilde{\textbf{w}}_{H}}= & \Big(\frac{1}{\tau_{H}^{2}} \textbf{I} + \frac{1}{\sigma_{H}^{2}}  \boldsymbol{\textbf{R}_{H,s}^{-1} \otimes \textbf{R}_{H,p}^{-1}}  \Big)^{-1} ,\label{eq:etywsywet}
\end{align}

The availability of $\tilde{\textbf{w}}_L$ and $\tilde{\textbf{w}}_H$ makes the computation of the missing variables easy since both $[\mathring{\boldsymbol{z}}_L|\tilde{\textbf{w}}_L,\boldsymbol{\beta}_L, \btau_L]$ and  $[\mathring{\textbf{z}}_L|\tilde{\textbf{w}}_L,\boldsymbol{\beta}_L, \btau_L]$ follow  independent normal distribution with constant diagonal variance. The conditional posterior $\pi(\bbeta,\rho,\bsigma^{2},\btau^2|\tilde{\textbf{z}},\tilde{\textbf{w}},\btheta)$ has the form:
\begin{align}
\boldsymbol{\beta}_{L} |\tilde{\textbf{z}},\tilde{\textbf{w}}, \bsigma^{2},\btheta  & \sim N \bigg(\hat{\bbeta}_{L}   , \hat{\textbf{V}}_{b,L}^{*} \bigg)\\
(\rho, \boldsymbol{\beta}_{H})|\tilde{\textbf{z}}, \tilde{\textbf{w}}, \bsigma^{2},\btheta, \bbeta_{L}  & \sim N \bigg( (\hat{\rho}, \hat{\bbeta}_{H}), \hat{\textbf{V}}_{b,H}^{*} \bigg)\\
\sigma_L^2 | \tilde{\textbf{w}}, \bsigma^{2},\btheta, \bbeta  & \sim IG \bigg(\hat{a}_{\sigma_L}, \hat{b}_{\sigma_L} \bigg)    \\
\sigma_H^2 | \tilde{\textbf{w}}, \bsigma^{2},\btheta, \bbeta    & \sim IG \bigg(\hat{a}_{\sigma_H},\hat{b}_{\sigma_H}\bigg)    \\
\tau_{L}^2 | \tilde{\textbf{z}}, \tilde{\textbf{w}}, \bbeta  & \sim IG \bigg(\hat{a}_{\tau_L}, \hat{b}_{\tau_L}\bigg)    \\
\tau_{H}^2 |\tilde{\textbf{z}}, \tilde{\textbf{w}}, \bbeta  & \sim IG \bigg(\hat{a}_{\tau_H},\hat{b}_{\tau_H} \bigg),   
\end{align}
where the hatted quantities are given in (19)-(30) of Appendix. 
The conditional posterior $p(\btheta_L| \tilde{\textbf{w}}_{L},\sigma_{L}^2)$ and $p(\btheta_H|\tilde{\bz}, \tilde{\textbf{w}},\sigma_{H}^2)$ cannot be sampled directly. Conditional independence in \eqref{eq:phiposterior-1} implies that $\btheta_L$ and $\btheta_{H}$ can be simulated by running in parallel two Metropolis-Hastings algorithms each of them
targeting distributions. These full conditional distributions are know up to some normalizing constant:
\begin{align*}
p(\boldsymbol{\theta}_{L}|\tilde{\textbf{w}}_{L},\sigma_{L}^2) & \propto  p(\boldsymbol{g}_{L}^{\boldsymbol{2}})p(\boldsymbol{\phi}_{L})\sigma_{L}^{-1}|\boldsymbol{\textbf{R}}_{L,s}|^{-1/2} |\textbf{R}_{L,p}|^{-1/2}\text{exp}\left\{ -\frac{1}{2\sigma_L^2}\tilde{\textbf{w}}_{L}^{T}(\boldsymbol{\textbf{R}_{L,s}^{-1} \otimes \textbf{R}_{L,p}^{-1}})\tilde{\textbf{w}}_{L}\right\}, \\
p(\boldsymbol{\theta}_{H}|\tilde{\textbf{w}}_{H},\sigma_{H}^2) & \propto  p(\boldsymbol{g}_{H}^{\boldsymbol{2}})p(\boldsymbol{\phi}_{H})\sigma_{H}^{-1}|\boldsymbol{\textbf{R}}_{H,s}|^{-1/2} |\textbf{R}_{H,p}|^{-1/2}\text{exp}\left\{ -\frac{1}{2\sigma_H^2}\tilde{\textbf{w}}_{H}^{T}(\boldsymbol{\textbf{R}_{H,s}^{-1} \otimes \textbf{R}_{H,p}^{-1}})\tilde{\textbf{w}}_{H}\right\}.
\end{align*}
In general the whole sampler can be described as a Metropolis-within-Gibbs sampler. This is a  computationally efficient sampler since avoids the inversion of big matrices and also most of the parameters are updated  via a closed form conditional distribution.

\subsection{Prediction}

Assume there is available a MCMC sample $\mathcal{S}^{N}=(\tilde{\textbf{w}},\boldsymbol{\sigma}^{\boldsymbol{2}},\boldsymbol{\theta},\btau^{2},\bbeta)$ generated from the MCMC sampler in Section \ref{Bayesian-inference}. The Central Limit Theorem can be applied to facilitate inference as the proposed sampler is aperiodic, irreducible, and reversible \citep{roberts2004general}.  The proposed latent variable co-kriging model with separable covariance structure (LVCS) allows inference to be performed for
the ``missing" output  $\mathring{\boldsymbol{y}}_{H}$
at input points in  $\mathring{\mathscr{D}}^{H}$.
Inference on  $\mathring{\boldsymbol{y}}_{H}$ can be particularly
useful when the retrieval algorithm has been unable to generate reliable output due to the presence of clouds ($QF=2$). The marginal
posterior distribution of $\mathring{\boldsymbol{y}}_{H}$, along
with its expectations, can be approximated via standard Monte Carlo
(MC) using the generated samples $\{ \mathring{\boldsymbol{y}}_{H}\}$. Alternatively, point estimates of $\mathring{\boldsymbol{y}}_{H}$
at $\mathring{\mathscr{D}}^{H}$ can be approximated by the more
accurate Rao-Blackwell MC estimator $\text{E}(\mathring{\boldsymbol{y}}_{H}|\boldsymbol{z}_{L},\boldsymbol{z}_{H})\approx\frac{1}{N}\sum_{j=1}^{N}\mathring{\boldsymbol{\mu}}_{H}^{(j)}$,
where $\{\mathring{\boldsymbol{\mu}}_{H}^{(j)}\}$ is the $j$-th mean
MCMC realization for $\mathring{\boldsymbol{y}}_{H}$.

To retrieve the temperature values $y_{H}(\mathscr{D}^{*})$ at unmeasured input points $\mathscr{D}^{*}$,  we propose a Monte Carlo recursive prediction procedure which is able to facilitate fully Bayesian predictive inference on the output. We first obtain $\boldsymbol{y}_{L}(\mathscr{D}^{*})$ and then use it to obtain  $\boldsymbol{y}_{H}(\mathscr{D}^{*})$. A direct prediction of  $y_{H}(\mathscr{D}^{*})$ can be computationally not feasible since the prediction distribution of the quality flag 0 cannot be simplified as it happens for the joint prediction distribution. The conditional distribution $[(\boldsymbol{y}_{L}(\cdot),\boldsymbol{y}_{H}(\cdot))|\tilde{\boldsymbol{w}}_{L},\tilde{\boldsymbol{w}}_{H}, \boldsymbol{\theta},\bsigma^{\boldsymbol{2}},\bbeta,\rho]$ inherits a conditional independence similar to the likelihood due to the augmentation of the latent variable $(\mathring{\boldsymbol{w}}_{L},\mathring{\boldsymbol{w}}_{H})$.  Hence, after integrating out $\bbeta,\rho$ and $\bsigma^{\boldsymbol{2}}$ we have  Student-T processes  (STP) for the conditional representation as: 
\begin{align}
y_{L}(\cdot)|\tilde{\boldsymbol{w}}_{L},\boldsymbol{\theta}_{L} & \sim \text{STP}\left(\boldsymbol{\mu}_{L}^{*}(\cdot|\tilde{\boldsymbol{w}}_{L},\boldsymbol{\theta}_{L}),\right.\hat{\sigma}_{L}^{2} \left.\hat{R}_{L}^{*}(\cdot,\cdot|\boldsymbol{\theta}_{L}),1+\tilde{n}\right);\label{eq:adhsdghh}\\
y_{H}(\cdot)|y_{L}(\cdot),\tilde{\boldsymbol{w}},\boldsymbol{\theta}_H & \sim \text{STP}\left(\mu_{H}^{*}(\cdot|\tilde{\boldsymbol{w}},\boldsymbol{\theta}_{H}),\right. \left.\hat{\sigma}_{H}^{2}\hat{R}_{H}^{*}(\cdot,\cdot|y_{L}(\cdot),\boldsymbol{\theta}_{H}),1+\tilde{n}\right),\label{eq:sghsfghf}
\end{align}
where 
\begin{align*}
\hat{\mu}_{L}^{*}(x|\tilde{\boldsymbol{w}}_{L},\boldsymbol{\theta}_L)= & \boldsymbol{H}_{L}(\boldsymbol{x};\tilde{\boldsymbol{w}}_{L})\hat{\boldsymbol{\beta}}_{L}+\boldsymbol{R}_{L}(x,\tilde{D})\boldsymbol{R}_{L}^{-1}(\tilde{D},\tilde{D})\tilde{\textbf{w}}_L\\
\hat{\mu}_{H}^{*}(x|\tilde{\boldsymbol{w}}_{H},\boldsymbol{\theta}_H)= & \hat{\mu}_{L}^{*}(x|\tilde{\boldsymbol{w}}_{L},\boldsymbol{\theta}_{L}) + \boldsymbol{H}_{H}(\boldsymbol{x};\tilde{\boldsymbol{w}}_{H})\hat{\boldsymbol{\beta}}_{H}+\boldsymbol{R}_{H}(x,\tilde{D})\boldsymbol{R}_{H}^{-1}(\tilde{D},\tilde{D})\tilde{\textbf{w}}_H
\end{align*}
and 
\begin{align*}
\hat{R}_{L}^{*}(\boldsymbol{x},\boldsymbol{x}'|\boldsymbol{\theta}_L)= & R_{L}(\boldsymbol{x},\boldsymbol{x}')-\boldsymbol{R}_{L}(\boldsymbol{x},\tilde{D})\boldsymbol{R}_{L}^{-1}(\tilde{D},\tilde{D})\boldsymbol{R}_{L}^{\top}(\boldsymbol{x}',\tilde{D})\\
 & \qquad+\left[\boldsymbol{H}_{L}(\boldsymbol{x})-\boldsymbol{R}_{L}(\boldsymbol{x},\tilde{D})\boldsymbol{R}_{L}^{-1}(\tilde{D},\tilde{D})\boldsymbol{H}_{L}(\tilde{D})\right]\hat{\boldsymbol{A}}_{L}\\
 & \qquad\qquad\times\left[\boldsymbol{H}_{L}(\boldsymbol{x}')-\boldsymbol{R}_{L}(\boldsymbol{x}',\tilde{D})\boldsymbol{R}_{L}^{-1}(\tilde{D},\tilde{D})\boldsymbol{H}_{L}(\tilde{D}_{L})\right]^{\top}\\
 \hat{R}_{H}^{*}(\boldsymbol{x},\boldsymbol{x}'|\boldsymbol{y}_{L},\boldsymbol{\theta}_H)= & R_{L}(\boldsymbol{x},\boldsymbol{x}')-\boldsymbol{R}_{H}(\boldsymbol{x},\tilde{D})\boldsymbol{R}_{L}^{-1}(\tilde{D},\tilde{D})\boldsymbol{R}_{H}^{\top}(\boldsymbol{x}',\tilde{D})\\
 & \qquad+\left[\boldsymbol{L}_{H}(\boldsymbol{x};\boldsymbol{y}_{L})-\boldsymbol{R}_{H}(\boldsymbol{x},\tilde{D})\boldsymbol{R}_{H}^{-1}(\tilde{D},\tilde{D})\boldsymbol{L}_{H}(\tilde{D};\boldsymbol{y}_{L})\right]\hat{\boldsymbol{A}}_{H}\\
 & \qquad\qquad\times\left[\boldsymbol{L}_{H}(\boldsymbol{x}';\boldsymbol{y}_{L})-\boldsymbol{R}_{H}(\boldsymbol{x}',\tilde{D})\boldsymbol{R}_{H}^{-1}(\tilde{D},\tilde{D})\boldsymbol{L}_{H}(\tilde{D}_{H};\boldsymbol{y}_{L})\right]^{\top}
\end{align*}
for $\boldsymbol{x},\boldsymbol{x}'\in\mathcal{\mathcal{X}}$, and  $\hat{\boldsymbol{A}}_{L}=(\boldsymbol{H}_{L}^T\boldsymbol{R}_{L}^{-1}(\tilde{D},\tilde{D})\boldsymbol{H}_{L})^{-1}$, $\boldsymbol{L}_{H}(\cdot;\boldsymbol{y}_{L})=\left[\boldsymbol{H}_{H}(\cdot),\boldsymbol{y}_{L}(\cdot))\right]$, $\hat{\boldsymbol{A}}_{H}=(\boldsymbol{L}_{H}^T\boldsymbol{R}_{H}^{-1}(\tilde{D},\tilde{D})\boldsymbol{L}_{H})^{-1}$,  $\boldsymbol{R}_{L}^{-1}(\tilde{D},\tilde{D})=\textbf{R}_{L,s} \otimes \textbf{R}_{L,p}$, $\boldsymbol{R}_{H}^{-1}(\tilde{D},\tilde{D})=\textbf{R}_{H,s} \otimes \textbf{R}_{H,p}$, $\boldsymbol{R}_{L}(\boldsymbol{x},\tilde{D})=(\textbf{R}_{L,s}(\bs,\tilde{D}) \otimes \textbf{R}_{L,p}(p,\tilde{D}))$, and $\boldsymbol{R}_{H}(\boldsymbol{x},\tilde{D})=(\textbf{R}_{H,s}(\bs,\tilde{D}) \otimes \textbf{R}_{H,p}(p,\tilde{D}))$.

The proposed prediction procedure integrates uncertainty regarding the unknown `missing data' and parameters. It is computationally preferable compared to a one step prediction, based on the prediction distribution of the  high fidelity only, because it allows the parallel inversion of smaller covariance matrices
with sizes $n_{s}\times n_{s}$ and $n_{p}\times n_{p}$ while the others
require the inversion of a large covariance matrix of size $n_{s}n_{p}\times n_{s}n_{p}$ .
Moreover, it is able to recover the whole predictive distribution and its moments. 

\section{Data Analysis and Results}\label{sec: res}

This section conducts a full analysis of the AIRS data set described in Section~\ref{sec: data} using the proposed latent variable co-kriging model with separable covariance structure (LVCS). We compare LVCS with two alternatives that ignores the flag values: (a) the latent variable separable Gaussian process (SGP) model;  (b) the non-separable additive approximate Gaussian process (AAGP) \citet{Ma_Konomi_Kang2019Env}. For both SGP and AAGP, data with QF values 0 and 1 are combined, and fit with the corresponding model, respectively. Analysis is performed in MATLAB R2020a, on a computer with specifications (intelR i7-3770 3.4GHz Processor, RAM 8.00GB, MS Windows 64bit). In addition, we consider two variations for each of the three models: a)  fix the variance of the nugget effects (${\bf g}_{\cdot}^2$)  within the correlation functions based on a cross validation exploration and b) sample ${\bf g}_{\cdot}^2$  within the MCMC procedure as explained in section~\ref{Bayesian-inference}. 

Each granule has  $135,000$ estimated temperature values of three different quality flags. Since the vertical pressure levels for atmospheric variables increase in an exponential order, we take a logarithmic transformation of the pressure level values.  In each specific FOR, the temperature is taken vertically at $100$ distinct pressure level values.  Pressure level values with non available (NA) data are removed prior to the analysis. These pressure level values are below the surface pressure and correspond to structurally missing/undefined data due to physical constraints.  

\begin{figure}[htb!]
	\centering
	\subfloat[\label{fig:reala}]{\includegraphics[width=0.45\textwidth]{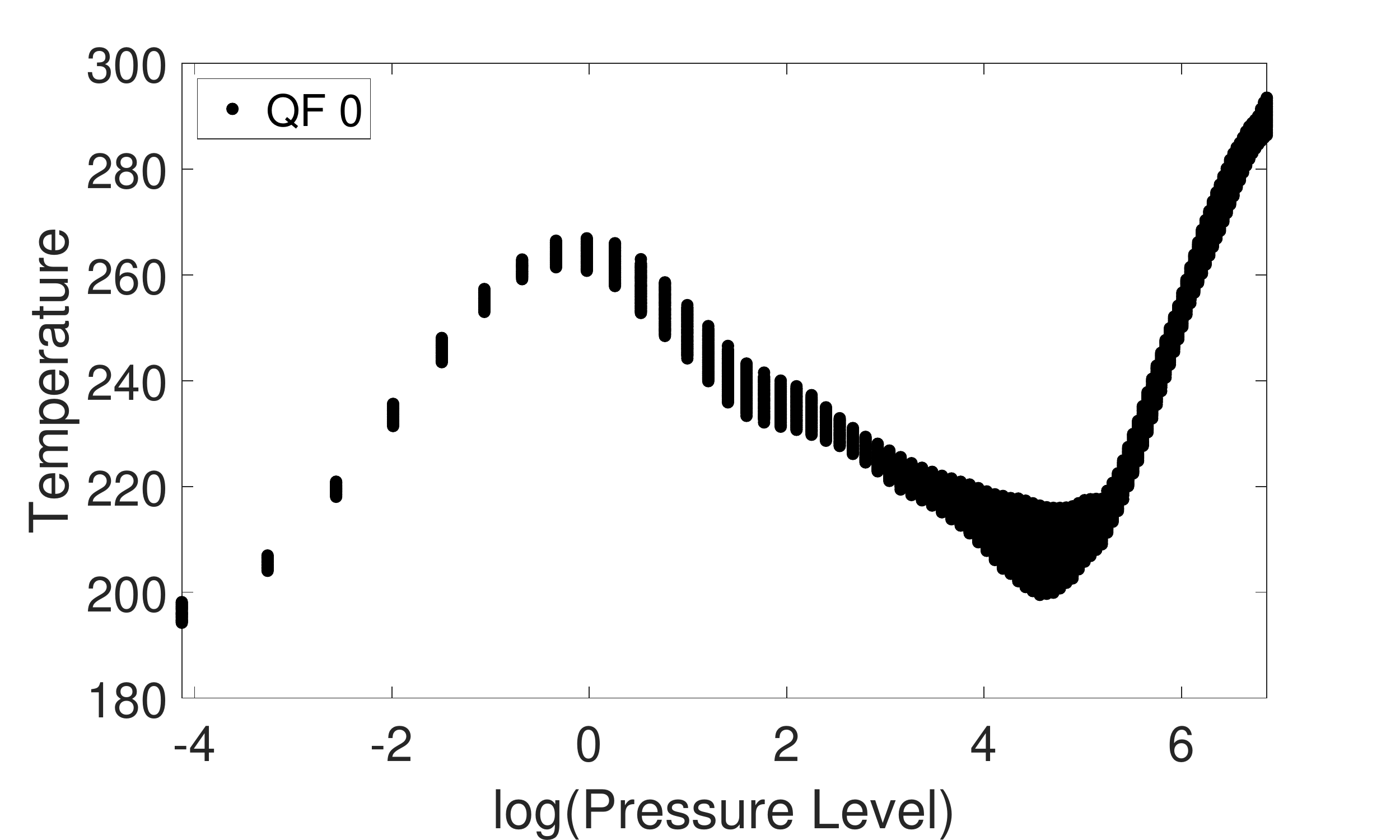}}
	\subfloat[\label{fig:realb}]{\includegraphics[width=0.45\textwidth]{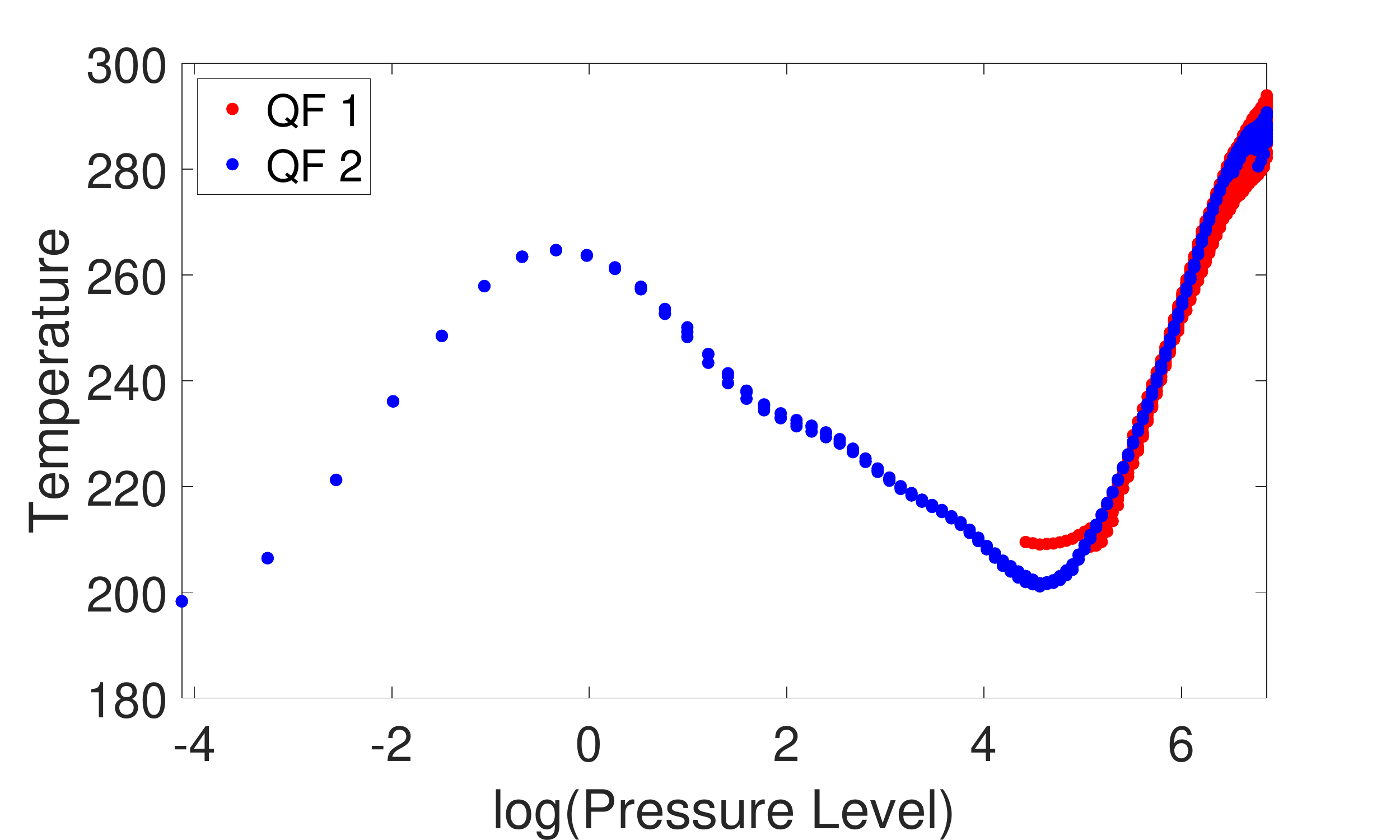}}
	\caption{AIRS retrieved temperature (units are Kelvin) as a function of log pressure levels: a) Quality flag 0 only b) Quality flag 1  (Red), and 2 (Blue).}	\label{fig:dataA}
\end{figure}

We start by testing our proposed model in a granule of August 1 of 2013 within the MAGIC validation campaign \citet{ZhouKollias2015}. The last 4 of the  distinct pressure levels have only NA data, so we consider only $96$ distinct pressure levels in our analysis.  Figure~\ref{fig:reala} shows the  temperature as a function of the log of pressure level values for a collection of FORs with only QF 0 and Figure~\ref{fig:realb} shows the temperature as a function of pressure level values for FORs with variable quality flags. From these plots as well as Figure~1, one can observe that the  QF 1 and 2 data  are more common in higher vertical pressure levels (closer to the sea surface) within the vertical profile. 

To simulate a realistic missing data (i.e., with QF 2) scenario, we randomly select testing data in a  block of input. Since missing observations are more common in higher vertical atmospheric pressure values (close to the sea surface), we only consider vertical atmospheric pressure values higher than $114.0070$ hPa (pressure level greater than $47$). Evaluation of predictive performance is based on mean squared prediction errors (MSPE), coverage probability of the $95\%$ equal tail credible interval (CVG($95\%$)), average length of the $95\%$ equal tail credible interval (ALCI($95\%$)), and continuous rank probability score (CRPS)\citep{GneitingRaftery2007}.


 Based on the  relationship of the log pressure values with the data Figure~\ref{fig:dataA}, we chose polynomial basis functions of degree three for the log pressure level. Since we do not observe a spatial structure of the mean of temperature given a pressure level, we chose a constant mean for the spatial basis. The basis functions of the complete data are created based on the Kronecker product of these two bases. Based on an empirical study between three different covariance functions; the square exponential, exponential, and the $5/2$ Mat\'ern; we chose the exponential covariance function.  For all the tested days and models, the exponential covariance function gave the best prediction performance. 

We apply the same specifications for all three models, the proposed LVCS, the separable model, and the AAGP. To speed up computations in the AAGP we specify a grid of knots: $15$ knots for pressure level, $8$ knots for longitude and $6$ for latitude. In total, we use $720$ knots for the predictive process part and the same specification for the separable part.  For the Bayesian inference of the three models on the unknown parameters $\bbeta_{1}, \bbeta_{2}$, we assigned independent Normal prior distributions with zero mean and large variances. We used inverse Gamma priors for the spatial and noise  variances  $\sigma_{L}^{2},\sigma_{H}^{2}, \tau_{L}^{2}, \tau_{H}^{2}$ as $IG(2,1)$. The range correlation parameters in space are assigned a uniform prior $U(0,100)$ and the range correlation parameters in logarithmic pressure level values a uniform prior $U(0,20)$ .  For all three models, we ran the MCMC sampler as described in Section $3$ with $25,000$ iterations where the first $5,000$ iterations were discarded as burn-in. For all the parameters of the three models the MCMC converges within the first $2,000$ iterations. The convergence of the MCMC sampler for each parameter was assessed from their associated trace plots.

Table 1 shows the results with the prediction performances for each model when we assume an unknown  nugget effect ($g_{\cdot}^2$)  within the correlation functions as described in section~3 and when we fix it to $10^{-6}$. Based on the MSPE and CRPS the proposed LVCS kriging gives better prediction results than the separable kriging and the AAGP for both fixed or random nugget effects cases. Making the nugget effect random improves the predictability for all models.   The proposed LVCS has smaller MSPE and CRPS than both the  separable kriging and the AAGP kriging. Specifically, the MSPE of the proposed model is less than half of both AAGP and separable model.  The LVCS has nominal coverage probability  close to $0.95$  and short interval  length for the  95\% credible interval.  The separable model and the AAGP kriging have very similar MSPE which indicates that AIRS data have a separable covariance structure. The computational time for the AAGP is almost $10$ times slower. It is worth pointing out that we can improve the computational time of AAGP by decreasing the number of predictive process knots. However, this may result in missing a potential non-separable small scale variation.

\begin{figure}[h]
	\centering
	\subfloat[\label{fig10:a}]{
		\includegraphics[width=7cm]{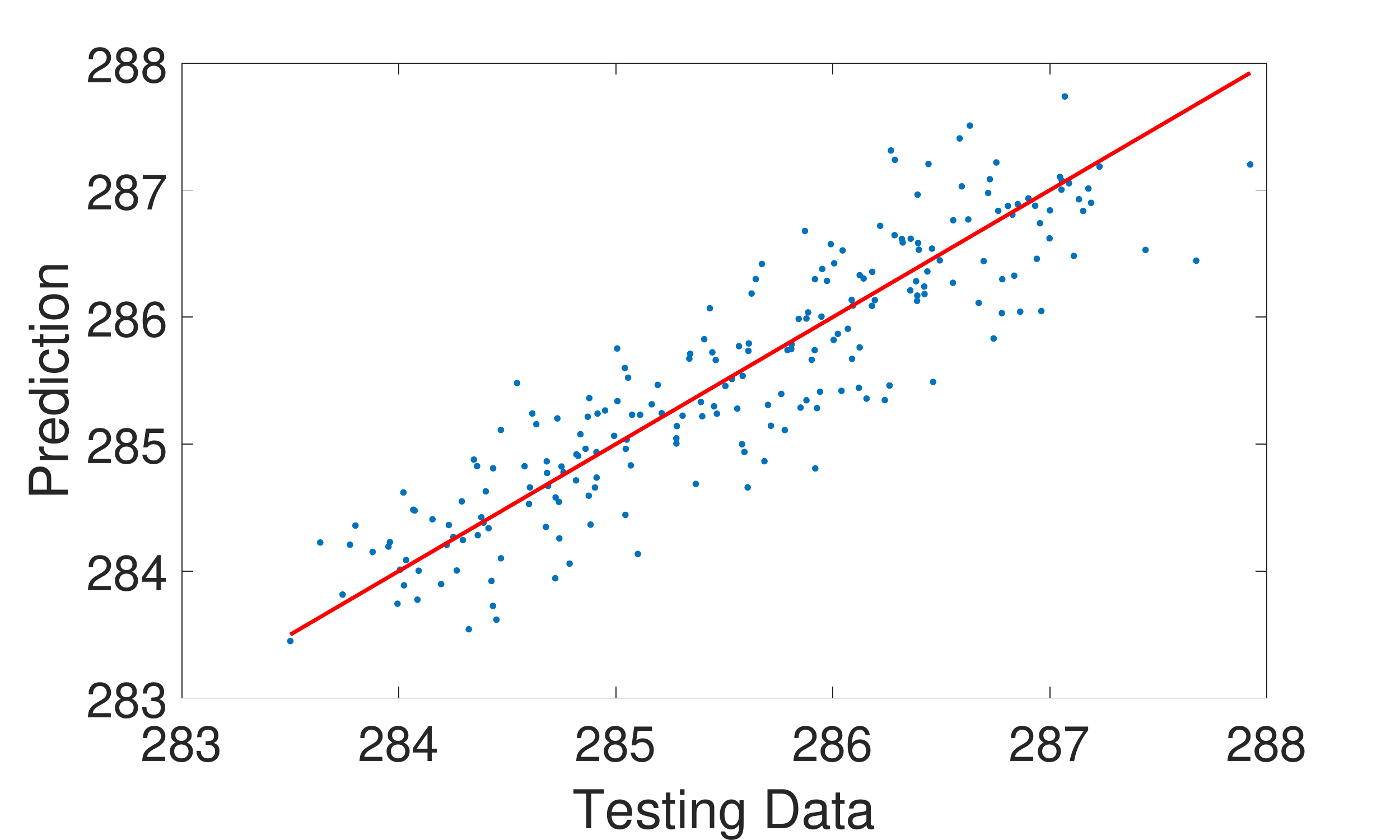}
	} \hspace{0.1in}
	\subfloat[\label{fig10:b}]{\includegraphics[width=7cm]{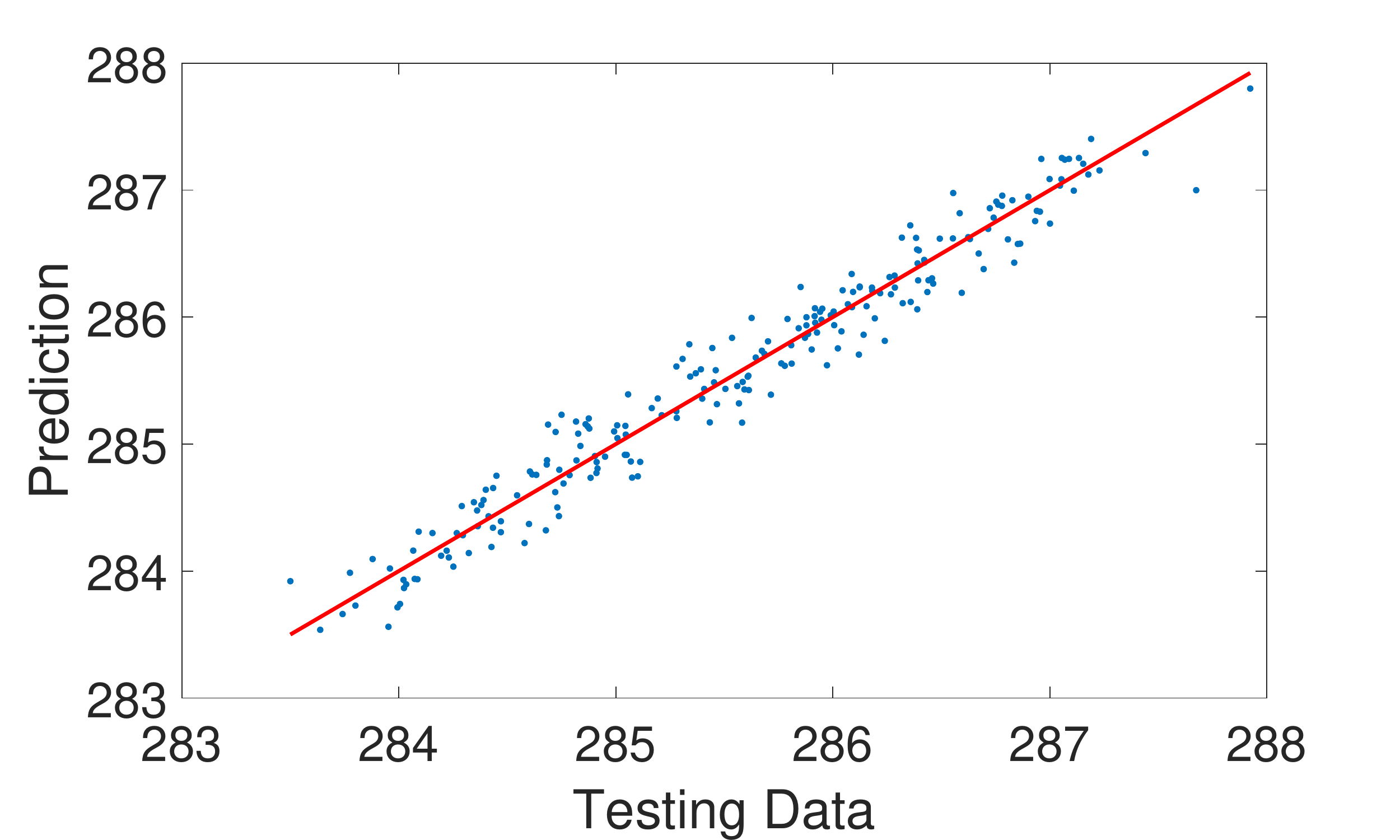}
	}
	\caption{Scatter plot of predicted values against held-out testing data for the last three pressure levels values using two different models: a) Separable model, and b) the proposed latent variable co-kriging model with separable covariance structure (LVCS)}
	\label{fig:10}
\end{figure}

To better demonstrate the prediction benefit of accounting for the quality flag into our model, we plot the predicted temperature values against held-out temperature values  for pressure level values between $750$ hPa and $850$ hPa using: (a)  Figure~\ref{fig10:a} the model which ignores the quality flag of the data and (b) Figure~\ref{fig10:b} the proposed LVCS model which accounts for different quality flags.  From these scatter plots it is clear that accounting for the different quality of the retrievals improves predictions. Specifically the  predicted  temperature values  of the proposed LVCS scattered in a narrower interval around  the  $45$  degree  straight red  line. The fact that the predicted values when ignoring the fidelity level are still scattered around the $45$  degree  straight red  line indicates that the flag one observations may be unbiased estimates of the temperature with bigger variance. 

\begin{figure}[h]
	\centering
	\subfloat[\label{fig:db}]{\includegraphics[width=0.45\textwidth]{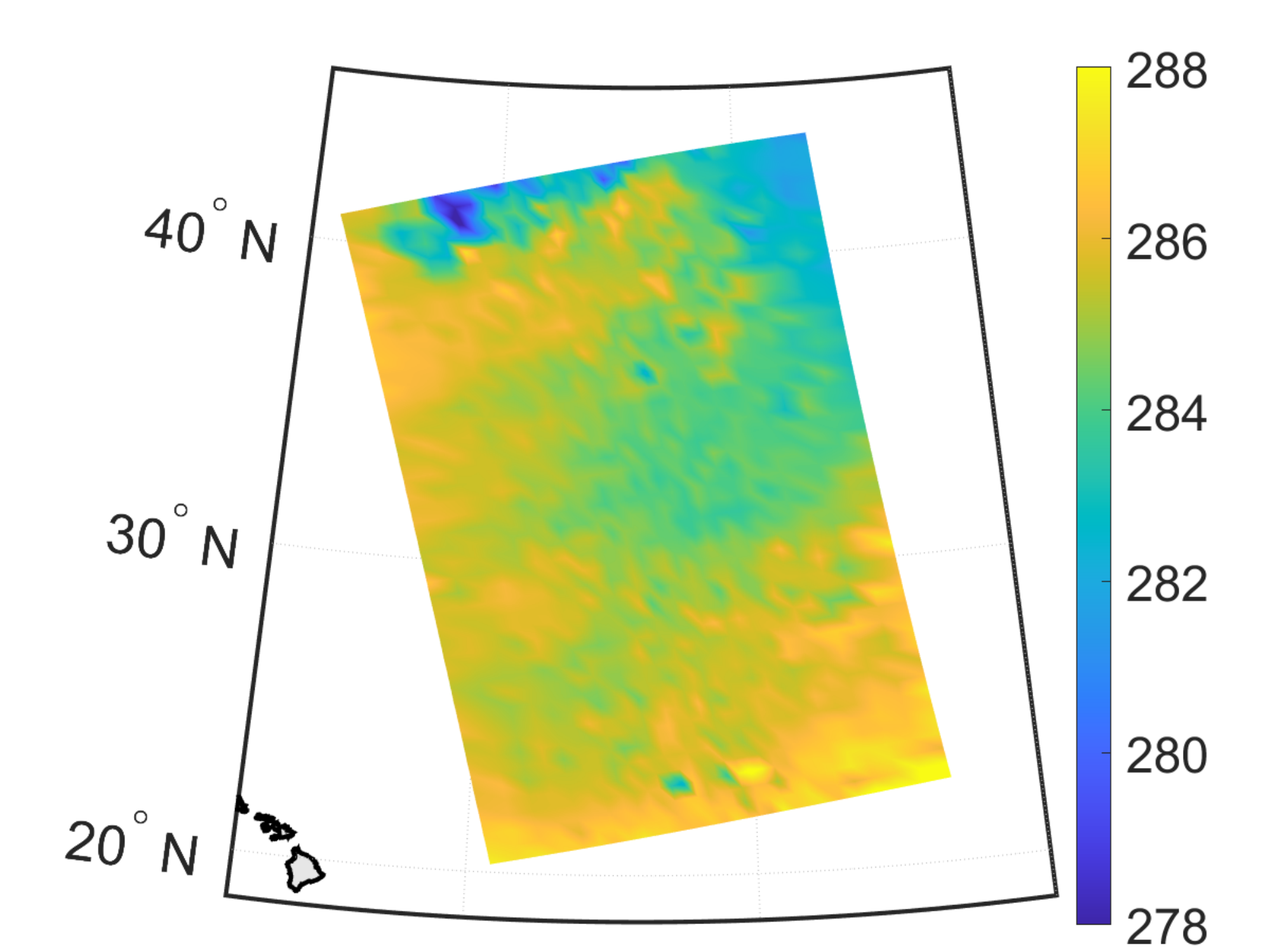}}
	\subfloat[\label{fig:de}]{\includegraphics[width=0.45\textwidth]{figures/RealTAirSupQCM_90_plot}}\\
	\subfloat[\label{fig:da}]{\includegraphics[width=0.45\textwidth]{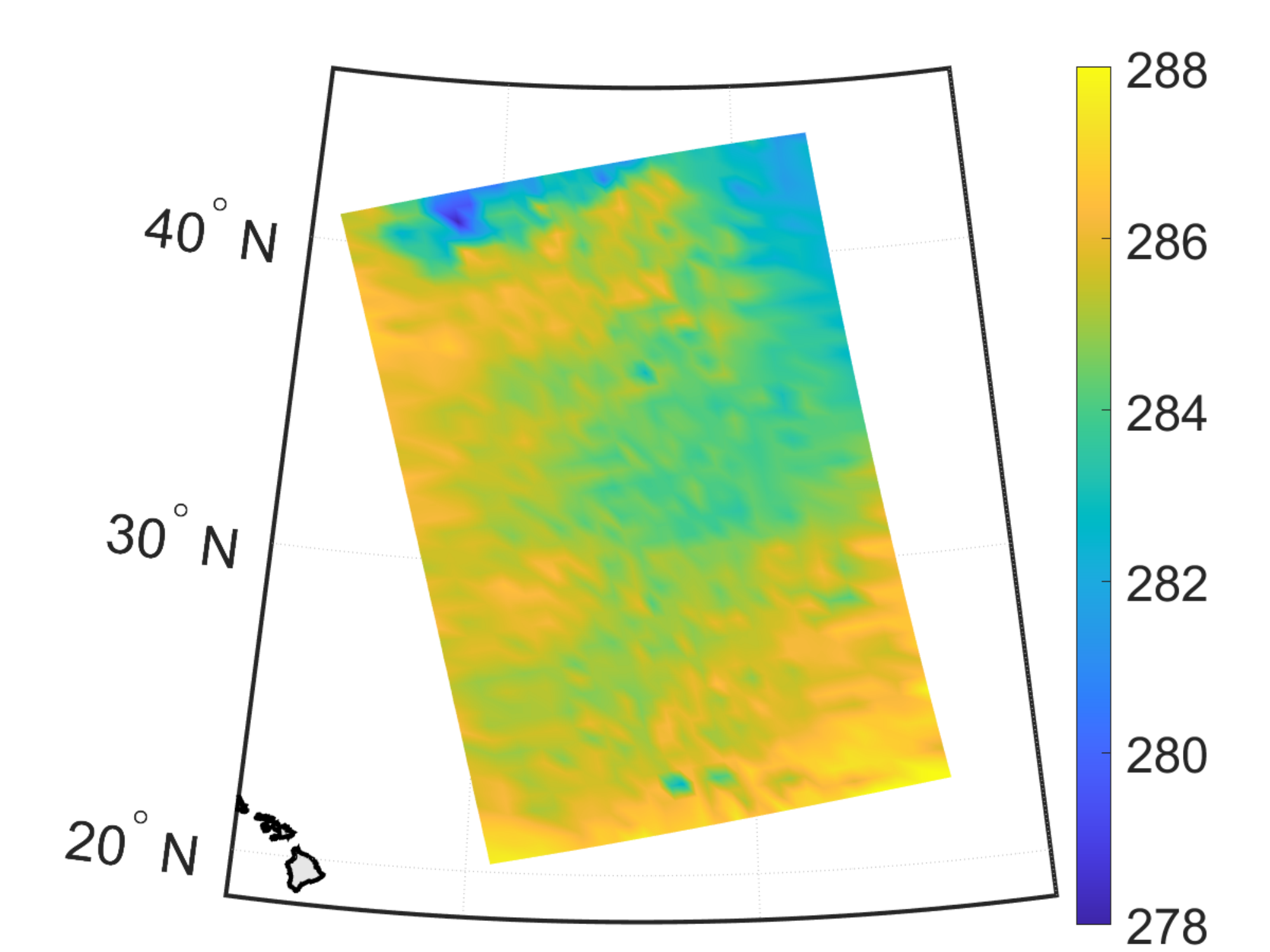}}
	\subfloat[\label{fig:dd}]{\includegraphics[width=0.45\textwidth]{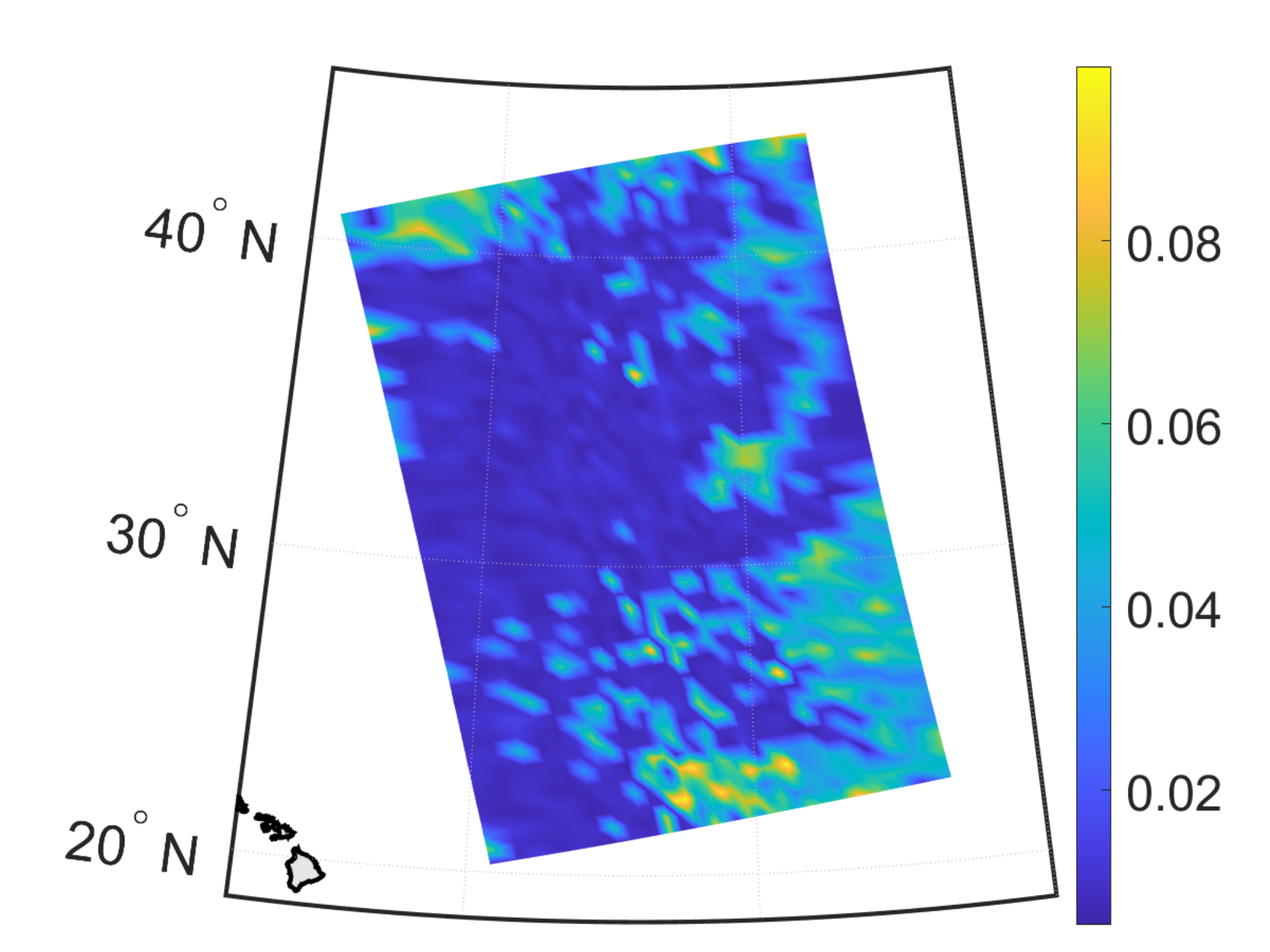}}
	\label{fig:data}
	\caption{a) AIRS air temperature data (units in Kelvin) b) the corresponding quality flag values c) Predicted means of the  temperature with quality flag value 0, d) Prediction standard error}
\end{figure}

We are able to reconstruct the 3D observations with the high fidelity (quality flag 0). Given the MCMC sample from fitting the model, the reconstruction based on the quality flag 0 data is automatic and does not require additional computational burden.  Figure ~\ref{fig:db} shows the observed temperature values of mixed quality flags at vertical pressure level $827$ hPa (pressure level 89). Figure~\ref{fig:de} shows the corresponding quality flag values of the data based on the locations.
Figure ~\ref{fig:da} shows the quality flag $0$ predicted mean values of the temperature at vertical pressure level $827$ hPa using the proposed recursive prediction procedure. The two images have a lot of similarities with more significant differences when the quality flag values are equal to 2 (Figure~\ref{fig:de}).  Also the proposed procedure can take into account the parameter uncertainty  as well as the nugget effects. Figure ~\ref{fig:dd} shows the predictive variance of the predicted values. There is a clear association of the the variance with the quality flag values and the spatial proximity as they are shown in Figure~\ref{fig:de}. 

\begin{figure}[h]
	\centering
	\subfloat[\label{fig:daa}]{\includegraphics[width=0.45\textwidth]{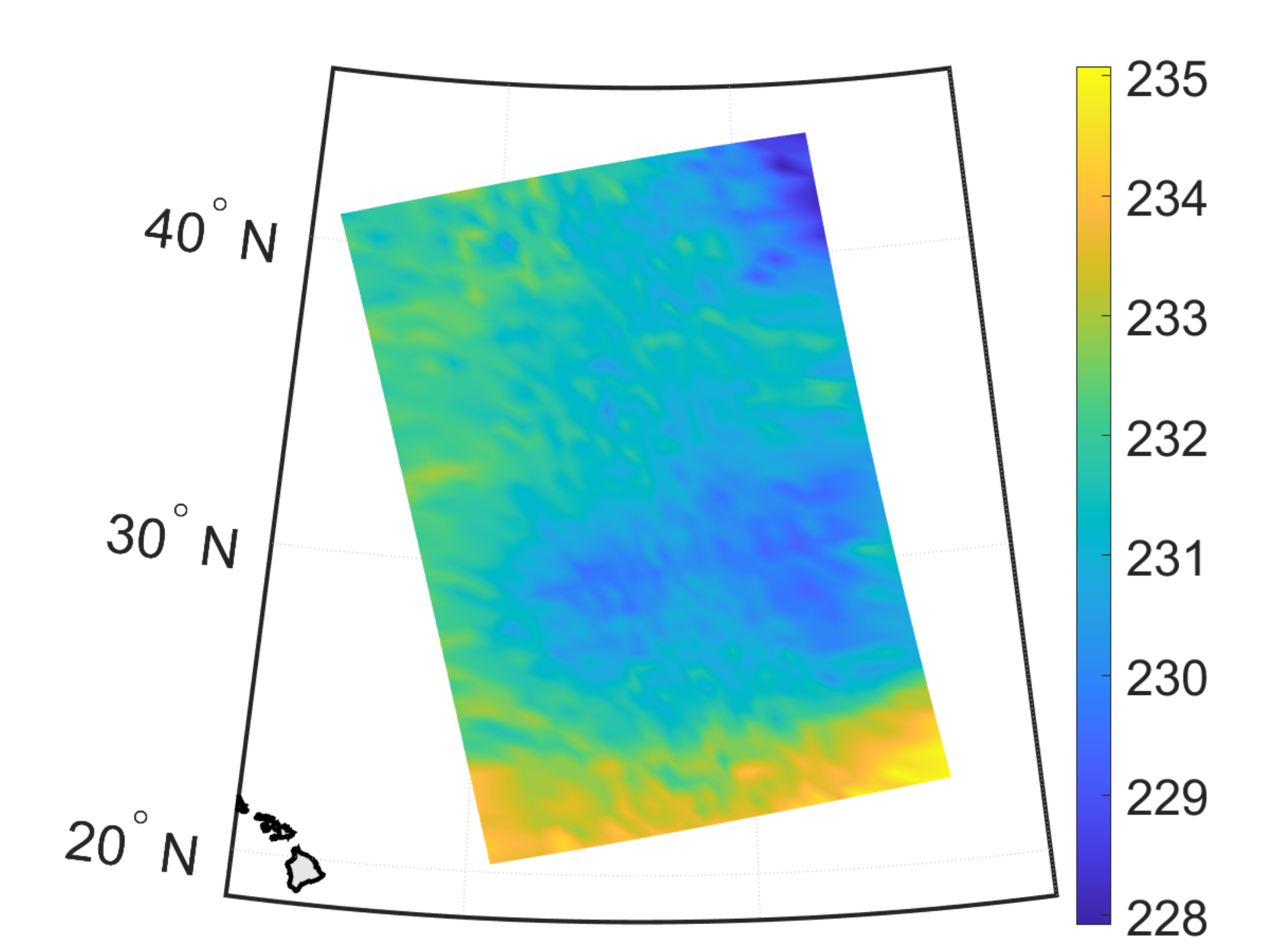}}
	\subfloat[\label{fig:dab}]{\includegraphics[width=0.45\textwidth]{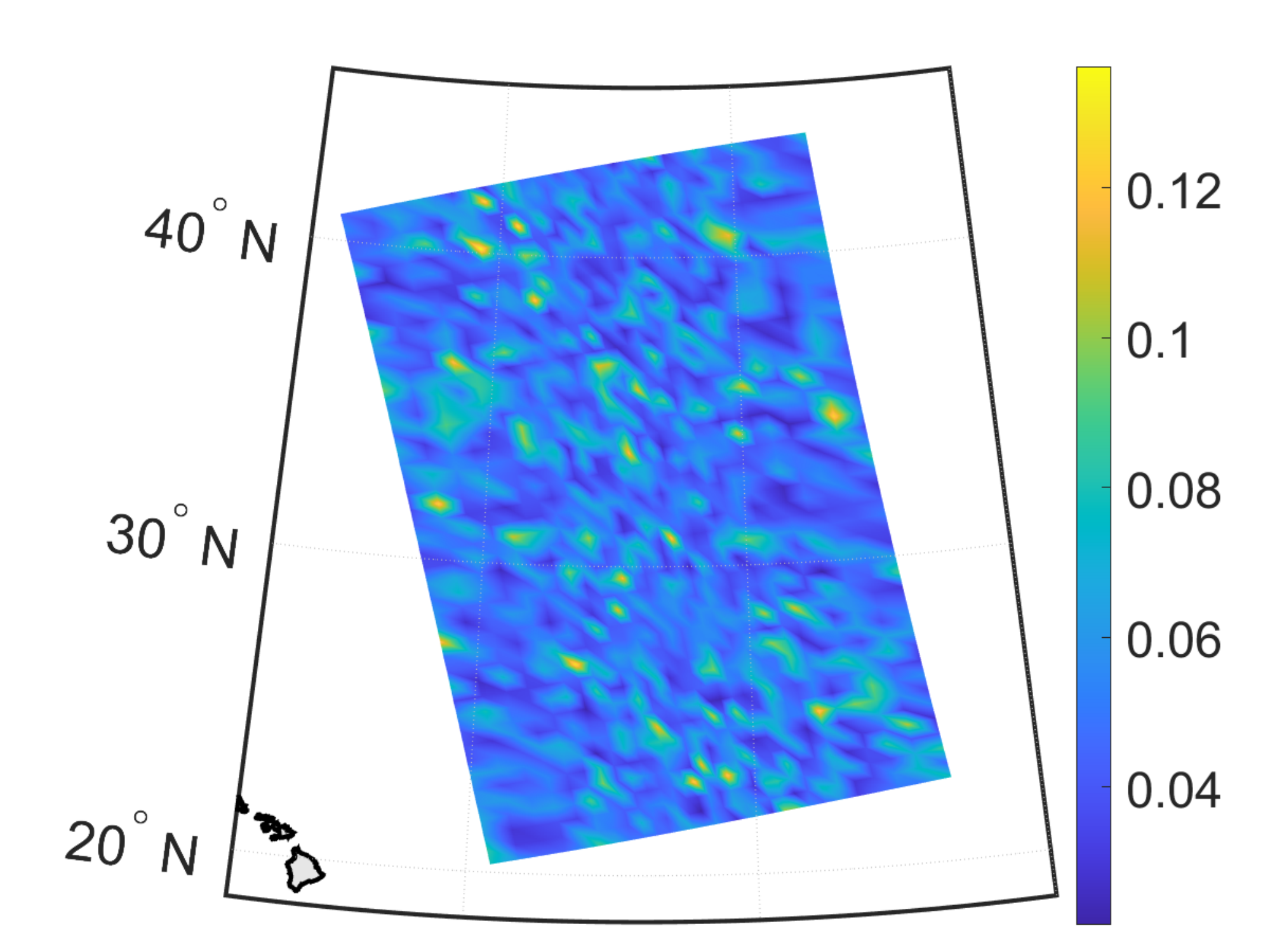}}
	\caption{a) Predicted means of the temperature (units are Kelvin) at $265$ hPa with quality flag value 0, b) Prediction standard error.}
	\label{fig:predic60} 
\end{figure}

Within the boundaries of a granule, we can also use the proposed procedure to predict the air temperature values at a unmeasured vertical pressure level different from those 100 levels in AIRS data products, i.e, interpolating vertically and horizontally. Given the MCMC values obtained in fitting of the model, we can find the predictive distribution for any location and vertical pressure value. With $20,000$ posterior samples of model parameters, the total computing time  to  make  prediction  for  a vertical pressure level at the specified granule locations  is  about  145.5 seconds based on Matlab 2020 on a laptop with  specifications described above. For example Figure~\ref{fig:daa} shows the predicted temperature and  Figure~\ref{fig:dab} shows the associated standard deviation for unmeasured vertical pressure level value $265$ hPa which corresponds to pressure level between $60$ and $61$. 

Within the study region of the subtropical eastern  Pacific Ocean,  we repeat the proposed model and inference for five different granules on different days in August 2013. Two different models are considered: a) the SGP which  ignores the quality flag and b) the proposed  LVCS which accounts for the quality flag. Because the AAGP gives similar results to the SGP and it is computationally more expensive, we discard it from this analysis. This is also a strong evidence of  the separable covariance structure between coordinates and atmospheric pressure  for the AIRS observation. We follow the same computational strategy as explained above where the MCMC is run in a sequence for both models. Table~\ref{table:MultipleDates} shows that the MSPE applying the LVCS is consistently almost half to the MSPE if we ignore the different flag quality. These results strengthen our modeling approach to account for different quality flags when we analyse AIRS temperature data sets.

\section{Conclusions and Discussion}\label{sec: conc}

In this paper, we propose a latent variable co-kriging model with separable Gaussian processes (LVCS) to analyze 3D AIRS air temperature data with different quality flags from NASA's AIRS instrument.  We propose an MCMC procedure with an imputation mechanism  which takes advantage of the 3D input structure to facilitate efficient computation. By applying our method to the AIRS data, we demonstrate that incorporating the quality-flag information into statistical modeling and data analysis can provide substantial inferential benefits. Unlike other methods for NASA's AIRS mission  in the literature \citep[e.g.,][]{AIRSV7Lev3,waliser_obs4mips}, we allow for different fidelity and missing data. Our methodology provides a coherent framework for the combined use of remote sensing retrievals with variable quality in both the horizontal and vertical directions. This capability can bolster the utility of observations from AIRS and next-generation infrared sounder instruments, particularly in challenging observing conditions such as the subtropical ocean regions illustrated in this work.

Our LVCS model can be generalized for larger spatial data sets over multiple granules. For example, we may apply a computationally efficient method such as low-dimensional conditional approximation \citep[e.g.,][]{datta2016hierarchical,katzfuss2017general} for both horizontal and vertical components in the separable covariance function. It is also possible to extend our method for multi-fidelity spatio-temporal remote sensing data by adding the temporal dependence and considering 4D data (longitude, latitude, vertical pressure level and time). Extending the LVCS modeling framework for multivariate multi-fidelity data could also prove to be fruitful. 

The current LVCS model assumes a linear relationship between low- and high-fidelity levels. One direction in future research is to extend this work by borrowing the idea of deep Gaussian process in \cite{perdikaris2017nonlinear} and  \cite{ming2021deep} so that we can incorporate nonlinear  dependence structure into the model. In addition, the LVCS modeling framework can be applied in observing system uncertainty experiments \citep{HobbsBraverman2017,Turmon2019,Braverman2021,Ma2021}, in particular when surrogates with different fidelity levels and computational efficiency are used to perform forward uncertainty propagation and inverse calibration for remote sensing data products.

\section{{Appendix}}
\subsection*{{Appendix A:} Direct Bayesian Inference\label{sec:Appendix}}


If we observe $\bZ=\{\bZ_{H},\bZ_{L}\}$ with their corresponding input $\bD=\{\bD^{H},\bD^{L}\}$. We can represent their likelihood as normal with mean:
\[E(\bZ)=\bH\bbeta
        = \begin{bmatrix}
       \bH_{L}(\bD^{H}) & \bH_{H}(\bD^{H})\\[0.3em]
        \bH_{L}(\bD^{L})  &  \bO 
       \end{bmatrix} \begin{bmatrix} \bbeta_{L} \\[0.3em] \bbeta_{H} \end{bmatrix},\]
 and its covariance matrix is: 
    \begin{equation*}
		\begin{split}
		\bV({\bZ})=\cov(\bZ,\bZ) &=
      \begin{pmatrix}
       \rho^2\bV_{L}(\bD^{H})+\bV_{H}(\bD^{H}) +\tau^2_{H}\bI_{n}  & \Cov_{L}(\bD^{H},\bD^{L}) \\[0.3em]
        \Cov_{L}(\bD^{L},\bD^{H})  &  \bV_{L}(\bD^{H})+ \tau^2_{L}\bI_{m}  
       \end{pmatrix},
    \end{split}
		\end{equation*}

The full joint posterior distribution of the GP hyperparameters $(\bTheta|\bZ)$ is  analytically intractable.
Exact posterior inference can be performed by a customized MCMC algorithm. Analytically, we firstly  we sample from the posterior distribution of $\bbeta,\bphi,\bsigma^2, \btau^2|\bZ$ with Metropolis-Hastings (M-H). Given the prior specification for $\bbeta$ and $\bsigma^2$ in Section \ref{sec: model}, the close form of the posterior distribution of $\bbeta$ given $\bphi,\btau^2,\bZ$ is a multivaraite Normal distribution with mean $\hat{\bbeta}=\bW\bH^{T}\bV_{\bZ}^{-1}\bZ$ and variance $\bW=(\bH^{T}\bV_{\bZ}^{-1}\bH)^{-1}$ is:
\begin{equation}\label{eq:posterior121} 
\bbeta| \bphi,\btau^2,\bZ \sim \calN(\hat{\bbeta}, \bW).
\end{equation}
Both $\hat{\bbeta}$ and $\bW$ depend on $(\bphi, \btau^2)$.  Using properties of the normal density function, we integrate out $\bbeta$  and compute the joint posterior distribution of $\bphi,\bsigma^2,\btau^2|\bZ$ as:
\begin{equation}\label{eq:posterior12}
\begin{split}
p(\bphi,\bsigma^2,\btau^2|\bZ) &\propto \pi(\bphi)\pi(\bsigma)\pi(\btau^2)|\bV_{\bZ}|^{-1/2}|\bW|^{1/2} \exp[-\frac{1}{2}(\bZ-\bH\hat{\bbeta})^T\bV_{\bZ}^{-1}(\bZ-\bH\hat{\bbeta})].
\end{split}
\end{equation}
 Non of the conditional posterior distributions of $\bphi|\bsigma^2,\btau^2$, $\bsigma^2|\bphi, \btau^2$ and  $\btau^2|\bsigma^2, \bphi$ can be sampled directly. Therefore, we can use Metropolis-Hastings updates \citep{Hastings1970} for the joint posterior or Metropolis-Hastings updates within a Gibbs sampler as in \citep{Mueller1993, Gelfand1990}. Another obstacle for the AIRS data is the computational complexity of the posterior. Based on the above representation, we need to  invert a covariance matrix of dimension $n_sn_p\times n_sn_p=135,000\times 135,000$  is practically impossible. 

\subsection*{{Appendix B:} Posterior specifications}\label{AppendixB}
Let $\textbf{G}_{H}=(\tilde{\by}_{L},\textbf{H}_{H})$
\begin{align}
\small
\hat{\textbf{V}}_{b,L}^{*}  & =\bigg(\tau_{L}^{-2}\textbf{H}_L^T\textbf{H}_L^T + \tau_H^{-2}\textbf{H}_{L}^T\textbf{H}_{L}+ \textbf{V}_{b,L}\bigg)^{-1}\\
\hat{\textbf{V}}_{b,H}^{*}  & =\bigg(\tau_H^{-2}\textbf{G}_{H}^T\textbf{G}_{H}+ \textbf{V}_{b,H}\bigg)^{-1}\\
(\hat{\rho},\hat{\bbeta}_{H}) & =\hat{\textbf{V}}_{b,L}^{*} \left(\tau_H^{-2}\textbf{G}_{H}^T(\tilde{\textbf{z}}_{H}-\tilde{\textbf{w}}_H) \right)\\
\hat{\bbeta}_{L} & =\hat{\textbf{V}}_{b,L}^{*} \left(\tau_{L}^{-2}\textbf{H}_L^T(\tilde{\textbf{z}}_{L}-\tilde{\textbf{w}}_L) + \tau_H^{-2}\textbf{H}_{L}^T(\rho^{-1}(\tilde{\textbf{z}}_{H}-\textbf{H}_{H}\boldsymbol{\beta}_{H}-\tilde{\textbf{w}}_H)-\tilde{\textbf{w}}_L) +\hat{\textbf{V}}_{b,L}\boldsymbol{\mu}_{b,L}\right)\\
\hat{a}_{\sigma_L}& =\frac{n}{2} + a_{\sigma_L}\\
\hat{a}_{\sigma_H}& =\frac{n}{2} + a_{\sigma_H}\\
\hat{b}_{\sigma_L}& = b_{\sigma_L} + \frac{1}{2} \tilde{\textbf{w}}_L^T  ( \boldsymbol{\textbf{R}_{L,s}^{-1} \otimes \textbf{R}_{L,p}^{-1}} )  \tilde{\textbf{w}}_L\\
\hat{b}_{\sigma_H} & = b_{\sigma_H} + \frac{1}{2} \tilde{\textbf{w}}_H^T  ( \boldsymbol{\textbf{R}_{H,s}^{-1} \otimes \textbf{R}_{H,p}^{-1}} ) \tilde{\textbf{w}}_H \\
\hat{a}_{\tau_L}& =\frac{n_{L}}{2} + a_{L,\tau}\\
\hat{b}_{\tau_L} & = b_{L,\tau} + \frac{1}{2} (\tilde{\textbf{z}}_{L}- \textbf{H}_{L} \boldsymbol{\beta}_{L}-\tilde{\textbf{w}}_L)^T   (\tilde{\textbf{z}}_{L}- \textbf{H}_{L} \boldsymbol{\beta}_{L}-\tilde{\textbf{w}}_L) \\
\hat{a}_{\tau_H}& =\frac{n_{H}}{2} + a_{H,\tau}\\
\hat{b}_{\tau_H}& =b_{H,\tau} + \frac{1}{2} (\tilde{\textbf{z}}_{H}- \textbf{H}_{L} \boldsymbol{\beta}_{L} - \textbf{H}_{H} \boldsymbol{\beta}_{H}-\tilde{\textbf{w}}_L -\tilde{\textbf{w}}_H)^T(\tilde{\textbf{z}}_{H}- \textbf{H}_{L} \boldsymbol{\beta}_{L} - \textbf{H}_{H} \boldsymbol{\beta}_{H}-\tilde{\textbf{w}}_L -\tilde{\textbf{w}}_H)
\end{align}

\newpage

\setstretch{1.5}
\singlespacing 

\section*{Acknowledgments}
The research of  Konomi and Kang  was partially supported by National Science Foundation grant NSF DMS-2053668. Kang was also partially supported by  Simons Foundation's Collaboration Award (\#317298 and \#712755), and the Taft Research Center at the University of Cincinnati.Part of this work was performed at the Jet Propulsion Laboratory, California Institute of Technology, under contract with NASA. Support was provided by the Atmospheric Infrared Sounder (AIRS) mission. The authors thank Amy Braverman and Hai Nguyen for valuable discussions during the course of this work.

\setstretch{1.5}
\singlespacing 
\setlength{\bibsep}{5pt}

 \bibliographystyle{jasa3}
\bibliography{reference1}

\newpage

\newpage


\newpage


\newpage


\newpage

\begin{table}[htbp]
\centering %
	\begin{tabular}{c c| c c c c c c}
		\hline\hline
		Model & Nugget & MSPE & CRSP  & CVG 95\% & ALCI 95\% & Time (hours) \tabularnewline
		\hline
		SGP & Fixed & 0.2714  & 0.3011  & 84.9\% & 1.8113 & 3.973 \tabularnewline 
		  & Random & 0.2503  & 0.2974   & 89.2\%  & 1.9213 & 5.977 \tabularnewline 
		\hline
		AAGP & Fixed & 0.2783 & 0.3361  & 91.7\% & 2.089 & 65.402
		\tabularnewline
		 & Random & 0.2482 & 0.3361  & 92.3\% & 2.139 & 96.419
		 \tabularnewline
		 \hline
		LVCS  & Fixed & 0.1341 & 0.2323  & 92.9\% &  1.58 & 7.801\tabularnewline
		    & Random  &  0.1081 & 0.2270  & 96.1\% & 1.63& 10.432\tabularnewline
		\hline
	\end{tabular}
\caption{Method comparisons.\label{table:SIMPLEexample}}
\end{table}

\begin{table}[htbp]
\centering %
	\begin{tabular}{c | c c c c c c c}
		\hline\hline
		Model & Aug. 1 & Aug. 4  & Aug. 6 & Aug. 8 & Aug. 9  \tabularnewline
		\hline
		Ignoring Quality Flag SGP & 0.2503  &  0.2963 & 0.6080  & 0.2668 & 0.2309  \tabularnewline 
		Proposed LVCS  &  0.1081 & 0.1603 & 0.2901  & 0.1203 &  0.0985  \tabularnewline
		\hline
	\end{tabular}
\caption{MSPE (in Kelvin) for multiple dates of two different models a) ignoring the quality flag and b) the proposed LVCS (which accounts for the quality flag). .\label{table:MultipleDates}}
\end{table}

\end{document}